\documentclass[showpacs,preprint,prb,aps]{revtex4}
\usepackage{amssymb}
\usepackage{epsfig}
\begin{document}
\title{Damping and decoherence of a nanomechanical resonator due to a few  two level systems }

\author{Laura G. Remus and Miles P. Blencowe}\affiliation{Department of Physics and Astronomy, Dartmouth College, Hanover, New Hampshire
03755, USA }
\author{Yukihiro Tanaka} \affiliation{Department of Applied Physics, Hokkaido University, Sapporo 060-8628, Japan}

\date{\today}

\begin{abstract}
We consider a quantum model of a nanomechanical flexing beam resonator interacting with a bath comprising a few damped tunneling two level systems (TLS's). In contrast with a resonator interacting bilinearly with an ohmic free oscillator bath (modeling clamping loss, for example), the mechanical resonator damping is amplitude dependent, while the decoherence of quantum superpositions of mechanical position states depends only weakly on their spatial separation.     
\end{abstract}


\maketitle

\section{\label{sec:introduction}Introduction}
The past few years have seen dramatic progress towards achieving the necessary conditions for demonstrating macroscopic quantum behavior in mechanical systems.\cite{aspelmeyer08}
Schemes under investigation typically consider either  micronscale mechanical resonators that are electrostatically coupled to superconducting qubits (quantum electromechanical systems)\cite{blencowe04,schwab05,naik06,lahaye09}
 or larger mechanical mirror resonators that couple via radiation pressure to light trapped in an optical cavity (optomechanical systems).\cite{aspelmeyer08}
One of the prime motivations for demonstrating quantum behavior is to deepen our understanding of the so-called quantum-classical divide, in particular how classical dynamics emerges from the underlying quantum dynamics as system sizes (i.e., mass/energy content) increase.\cite{schlosshauer07}
It is commonly accepted that the environmental degrees of freedom with which the mechanical resonator's mode of interest interacts is responsible for the emergence of classicality.\cite{zeh85,zurek91}
 In particular, the environment is thought to cause the rapid  decoherence of initial quantum superposition states of the mechanical mode, resulting in an apparent classical mixture of the states. There is a considerable body of theoretical work  investigating  the effective quantum dynamics of  open, single particle systems.\cite{caldeira83,weiss99}
However, largely for reasons of calculational convenience, much of the effort has been devoted to the solvable model of an environment comprising noninteracting oscillators that are bilinearly coupled to a single oscillator system.\cite{grabert88,paz01} 
In light of the experimental progress mentioned above, an important issue is the actual nature of the dominant mechanical resonator mode environments. At the very low (i.e., cryogenic) temperatures to which the resonators must be cooled in order to observe quantum effects, it is not a priori obvious that  the resonator mode dynamics can be mapped onto that of the oscillator system-oscillator bath model. 
In the present paper, we focus on a  type of environment degree of freedom that is known to be relevant at low temperatures, namely the tunneling two level system.\cite{phillips87,esquinazi98}

Tunneling two level system (TLS) defects were first invoked in the early seventies in order to account for the observed thermodynamic properties of amorphous, dielectric materials at low temperatures.\cite{phillips72,anderson72}
Further, convincing evidence for their presence was provided by acoustic phonon pulse decay and phonon pulse echo experiments.\cite{arnold75,golding76,graebner79} In particular, these experiments verified the characteristic saturation of TLS's with increasing acoustic pulse intensity for resonant phonon absorption and also measured TLS relaxation and dephasing times. 
TLS's have recently received renewed interest as one of the main decay/decoherence mechanisms for superconducting qubits.\cite{simmonds04,martinis05,shnirman05,ku05,tian07,oconnell08,neeley08,constantin09}
Signatures include inducing resonant splittings in the  qubit energy level spectra\cite{simmonds04,martinis05,neeley08} 
and saturation of microwave power absorption by the dielectric oxide layer of the qubit tunnel junctions.\cite{martinis05,oconnell08} 
In contrast with the bulk amorphous dielectric materials involved earlier investigations, the much smaller, micronscale sizes of the superconducting qubits  with  gigahertz frequency energy level separations exceeding the dilution fridge $k_B T$ thermal energies point to a distinct and less explored regime in which the system qubit resonantly couples strongly to only a few TLS defects, as opposed to a dense spectrum.
Similarly, it may be the case that, given the much smaller volumes ($\sim 10^{-18}~{\mathrm{m}}^3$) of the micronscale mechanical resonators currently under investigation, the relevant system-environment model is an oscillator interacting with only a few TLS's. In the following sections we shall analyze just such a system. 

In Sec.~\ref{sec:estimates} we give some simple estimates based on existing bulk system TLS theory  in order to motivate our  mechanical resonator mode-few TLS model as well as to anticipate some of the consequences of TLS-dominated mechanical damping/decoherence. In Sec.~\ref{sec:hamiltonian} we derive the model closed system resonator-TLS Hamiltonian, and in Sec.~\ref{sec:open} we present the open system master equation   with further details of the derivation given in the Appendix. Section~\ref{sec:damping} focuses on the effect of a few TLS's on the resonator damping, while Sec.~\ref{sec:decoherence} describes the consequences for the decoherence of mechanical resonator superposition states.  Section~\ref{sec:conclusion} provides a few concluding remarks.

Although the present paper focuses on the role of TLS's for micronscale mechanical oscillator damping/decoherence, we do not completely neglect other mechanisms. In the spirit of keeping our model as simple as possible, we lump together all other relevant damping and decoherence mechanisms, such as clamping loss,\cite{cross01,photiadis04,geller05,wilsonrae08}
as an additional oscillator bath to which the mechanical system mode couples. This will allow us to  gauge somewhat the extent to which  other baths `interfere'  with the TLS bath in their damping and decoherence effects on the oscillator system. For example, the system oscillator's net damping rate need not be the sum of the damping rates due to the individual baths. One point that should be emphasized in this context is the highly nonlinear, quantum nature of the  coupled oscillator-TLS (equivalently spin-$1/2$) dynamics. Exact analytical or even simpler approximate equations are hard to come by and so we will  resort to solving for the full dynamics using numerical methods. We will be limited computationally to considering only a few TLS's--three to be precise. In future work  we plan to find ways to analyze the effects on mechanical damping/decoherence of  larger numbers of TLS's.

Another source of nanomechanical damping and decoherence that we do not explicitly take into account is the measurement process itself.\cite{braginsky92,clerk04,clerk08} The  resonator damping described in Sec.~\ref{sec:damping} can  be probed using, for example, continuous in time position detection with a single electron transistor.\cite{blencowe04,naik06}  We shall assume that the resulting back reaction on the resonator due to the position detector  can be simply modeled by the same additional  oscillator bath  at some finite temperature.\cite{blencowe05,clerk05} The  resonator superposition state decoherence described in Sec.~\ref{sec:decoherence}  can be probed using, for example, the microwave cavity-superconducting qubit scheme outlined in Refs.~[\onlinecite{armour08,blencowe08}].        

In the  present paper we neglect mechanical strain (i.e., phonon) mediated coupling between TLS's, assuming the latter to couple directly only to the oscillator system mode and with the TLS's damping treated phenomenologically, characterized by a decay time $T_1$. Given our current, almost complete lack of theoretical understanding of the role of TLS's for damping and decoherence in nano-to-mesoscale mechanical resonators, we feel that it is worthwhile to start with this simpler, noninteracting TLS model. 
The low temperature acoustic pulse probe investigations of bulk amorphous solids\cite{arnold75,graebner79}
and the mechanical quality factor and resonant frequency measurements of much larger resonators\cite{classen99,fefferman08}
point to the likely importance of interactions between TLS's.\cite{black77,burin95,burin98}
For example, phonon echo experiments yield TLS dephasing times  that are much shorter than their $T_1$ lifetimes, thought to be due to TLS spectral diffusion arising from the phonon mediated interaction between non-resonant TLS's.\cite{black77}
With the significantly reduced volumes of nano-to-mesoscale mechanical resonators, these strain interactions may in fact be considerably enhanced. We  plan to analyze the effects of such TLS interactions on nanomechanical damping/decoherence in a future work.

\section{\label{sec:estimates}Some Estimates}
In this section we adopt various results from earlier analyses of bulk, amorphous systems to try to gain some initial idea of expected consequences for damping/decoherence of nano-to-mesoscale mechanical resonators due to TLS's. We begin by estimating the number magnitude of TLS's that are near resonance with the mechanical mode frequency of interest, which we shall in this paper assume to be the lowest, flexural mode. For a range of bulk, amorphous solids, experiments are consistent with a TLS distribution of the form\cite{phillips87}
\begin{equation}
dN=V\frac{\bar{P}}{\Delta_b}d\Delta_0 d\Delta_b,
\label{distributioneq}
\end{equation}
where $V$ is the mechanical resonator volume, $\Delta_0$ and $\Delta_b$ are the asymmetry and tunnel splitting energies of the TLS's potential double well (see Fig.~\ref{beamTLS}), and $\bar{P}$ is the approximately constant spectral density that can be expressed as
\begin{equation}
\bar{P}=C\frac{\rho v^2}{\nu^2}.      
\label{TLSspectraldensityeq}
\end{equation}
Here, $\rho$ is the mass density, $\nu$ is the deformation potential (approximated as isotropic) and $v$ is the speed of sound (approximated as isotropic and polarization independent).
The dimensionless constant $C\sim 10^{-4}{\mathrm{-}}10^{-3}$ is approximately universal.\cite{berret88,pohl02} Of course, the nanomechanical resonator may not be fashioned out of one of the amorphous materials surveyed in Ref.~[\onlinecite{pohl02}],
but instead out of a crystalline material. In such a case we view Eq.~(\ref{distributioneq}) and the estimates we shall now derive for the total TLS number as providing an upper bound. Integrating Eq.~(\ref{distributioneq}) to obtain the total number of TLS's in energy width $\delta E$ about the TLS eigenenergy $E=\sqrt{\Delta_0^2 +\Delta_b^2}$, we obtain:
\begin{equation}
\delta N=\frac{1}{2} V\bar{P}\ln\left[\frac{1+\sqrt{1-(E_{\mathrm{min}}/E)^2}}{1-\sqrt{1-(E_{\mathrm{min}}/E)^2}}\right]\delta E,
\label{N1eq}
\end{equation} 
where $E_{\mathrm{min}}$ is the low energy cutoff in the TLS distribution.\cite{lasjaunias78}
Taking the typical ballpark values $\rho v^2\sim10^{11}~{\mathrm{kgm}}^{-1}{\mathrm{s}}^{-2}$ and $\nu\sim 1~{\mathrm{eV}}$ gives for the spectral density (\ref{TLSspectraldensityeq}), $\bar{P}\sim 10^{44}{\mathrm{-}}10^{45}~{\mathrm{J}}^{-1}{\mathrm{m}}^{-3}$. Substituting this into Eq.~(\ref{N1eq}) and taking $E=\hbar\omega$ (i.e., in resonance with the oscillator mode $\omega$), $\delta E=\hbar\omega/Q$, where $Q$ is the oscillator mode quality factor, and $E_{\mathrm{min}}/k_B=1~{\mathrm{mK}}$, we obtain for our estimated total TLS number  close to resonance expressed in natural units:
\begin{equation}
\delta N\sim 0.1{\mathrm{-}}1 \frac{\left(V/\mu{\mathrm{m}}^3\right) \left(f/{\mathrm{GHz}}\right)}{\left(Q/10^4\right)}.
\label{N2eq}
\end{equation}
Thus, according to this estimate, micron-scale, radio frequency mechanical resonators are on the borderline between  being unlikely to have a single TLS close to resonance and being very likely with a moderate increase in size.

We next estimate  the nanomechanical fundamental mode displacement amplitudes  that saturate the TLS's. Observation of quantum effects in the mechanical dynamics will require cooling the resonator such that $k_B T\lesssim \hbar\omega$. In this regime resonant absorption damping is expected to dominate. Theory of resonant TLS mechanical damping and acoustic pulse decay in large amorphous resonators and bulk solids gives for the energy damping rate\cite{hunklinger76}
\begin{equation}
\Gamma_{\mathrm{damp}}\approx\frac{\pi C\omega}{\sqrt{1+(\epsilon\nu/\hbar)^2 T_1 T_2}}\tanh\left(\frac{\hbar\omega}{2 k_B T}\right),
\label{gammabulkeq}
\end{equation}
where $C$ is the same dimensionless constant as introduced above, $\epsilon$ is the elastic strain amplitude (approximated as isotropic), $\nu$ is the deformation potential, $T_1$ is the TLS damping time, and $T_2$ is the TLS transverse relaxation time, related to the dephasing time $T_{\phi}$ as $T_2^{-1}=(2 T_1)^{-1}+T^{-1}_{\phi} $. From this expression we see that the saturation threshold strain is given by
\begin{equation}
\epsilon\sim\frac{\hbar}{\nu\sqrt{T_1 T_2}}.
\label{saturationeq}
\end{equation}
For the example of a doubly-clamped beam mechanical resonator of length $l$, thickness $d$, and midpoint $l/2$ transverse displacement amplitude  $Y$ (see Sec.~\ref{sec:hamiltonian}), the volume averaged, rms strain is $\bar{\epsilon}\sim d Y/l^2$. Substituting this into Eq.~(\ref{saturationeq}) and expressing in terms of natural units, we obtain
\begin{equation}
\left(Y/{\mathrm{\AA}}\right)\sim 10^{-6}\frac{\left(l^2/\mu{\mathrm{m}}^2\right)}{\left(\nu/{\mathrm{eV}}\right)\left(d/\mu{\mathrm{m}}\right)\sqrt{\left(T_1 T_2/\mu{\mathrm{sec}}^2\right)}}.
\label{saturation2eq}
\end{equation}   
Earlier phonon echo experiments in fused silica glass found TLS damping and transverse relaxation times $T_1\sim 100~\mu{\mathrm{sec}}$ and $T_2\sim 10~\mu{\mathrm{sec}}$, respectively, at $T\approx 20~{\mathrm{mK}}$.\cite{graebner79}
 Thus, the transverse relaxation is dominated by dephasing: $T_2\approx T_{\phi}$. Given that quantum zero-point displacement uncertainties of micron-scale mechanical resonators are $\sim 10^{-4}{\mathrm{-}}10^{-3}~$\AA,\cite{blencowe04} which, as can be seen from Eq.~(\ref{saturation2eq}), exceed the saturation threshold, we thus expect that experiments which measure damping and decoherence of such resonators will operate well into the saturation regime. As can be seen from Eq.~(\ref{gammabulkeq}), one important consequence is that resonant TLS dominated damping is expected to be amplitude dependent, corresponding to a nonlinear damping force that depends on both the position and velocity coordinates of the mechanical oscillator.
 
 Our final approximation concerns the decoherence of mechanical superposition states. For weak mechanical damping (i.e., $Q\gg 1$) and provided temperatures are not too low (i.e., $k_B T\gg \hbar\omega/Q$), the effective dynamics of an oscillator bilinearly interacting with a bath of free oscillators satisfies a quantum fluctuation-dissipation relation between the system oscillator's damping and decoherence rates:
 \begin{equation}
 \Gamma_{\mathrm{decohere}}=\frac{1}{2}\Gamma_{\mathrm{damp}}\left(\frac{\Delta Y}{Y_{zp}}\right)^2 \coth\left(\frac{\hbar\omega}{2 k_B T}\right),
 \label{qfdeq}
 \end{equation}
 where $\Delta Y$ is the uncertainty in the oscillator's position and $Y_{zp}=\sqrt{\hbar/(2m\omega)}$ is the oscillator's quantum zeropoint position uncertainty. Assuming that this fluctuation-dissipation relation applies also to the bulk oscillator with the TLS bath, we can substitute in expression~(\ref{gammabulkeq}) for the damping rate to obtain the decoherence rate. The first thing to notice is the cancellation of the hyperbolic temperature functions. Below the saturation threshold we would therefore conclude that the decoherence rate is temperature {\emph{independent}} while the damping rate is temperature dependent for an oscillator coupled to a TLS bath. In this respect, the oscillator system-TLS bath is `dual' to the oscillator system-oscillator bath in the sense that the decoherence rate of the latter has the inverse temperature dependence of the damping rate of the former, while the damping rate of the latter and decoherence rate of the former are both temperature-independent.\cite{schlosshauer08}
 However, this duality is to a certain extent academic since the above saturation estimates suggest that  nano-to-mesoscale mechanical resonators will be well within the saturation regime for quantum superpositions of distinct position states that must necessarily be larger than the zeropoint position uncertainty. As a consequence, the temperature dependences of the TLS $T_1$ and $T_2$ dependent terms must also be taken into account. Since a good understanding of the relaxation mechanisms of TLS's  is  lacking in nano-to-mesoscale mechanical resonators,\cite{seoanez08} we will not attempt to make predictions for the temperature dependences of various observable quantities in the present paper. 
 
 Leaving aside temperature dependencies, another notable consequence of applying the above quantum fluctuation-dissipation relation to the oscillator system-TLS bath is the weaker (i.e., linear)  dependence of the decoherence rate on oscillator position uncertainty. Thus, at low temperatures we might expect the decoherence rate to increase more gradually as the position separation in the quantum superposition state is increased, as compared with the quadratic separation dependence for the oscillator bath. Of course, other decoherence mechanisms, e.g, due to clamping loss, will then be expected to eventually dominate if they have the stronger quadratic dependence. 
 
 Summarizing the findings of this section, we expect that a relevant model for a nano-to-micronscale mechanical resonator interacting with TLS's is a single oscillator coupled to a bath of a few TLS's. Furthermore, we expect that mechanical damping will show an amplitude dependence, while the decoherence of mechanical superposition states  will depend weakly on their position separation. The following sections will bear out these expectations.

\section{\label{sec:hamiltonian}Resonator-TLS Hamiltonian}
\begin{figure}[htbp]
	\centering
		\includegraphics[width=4in]{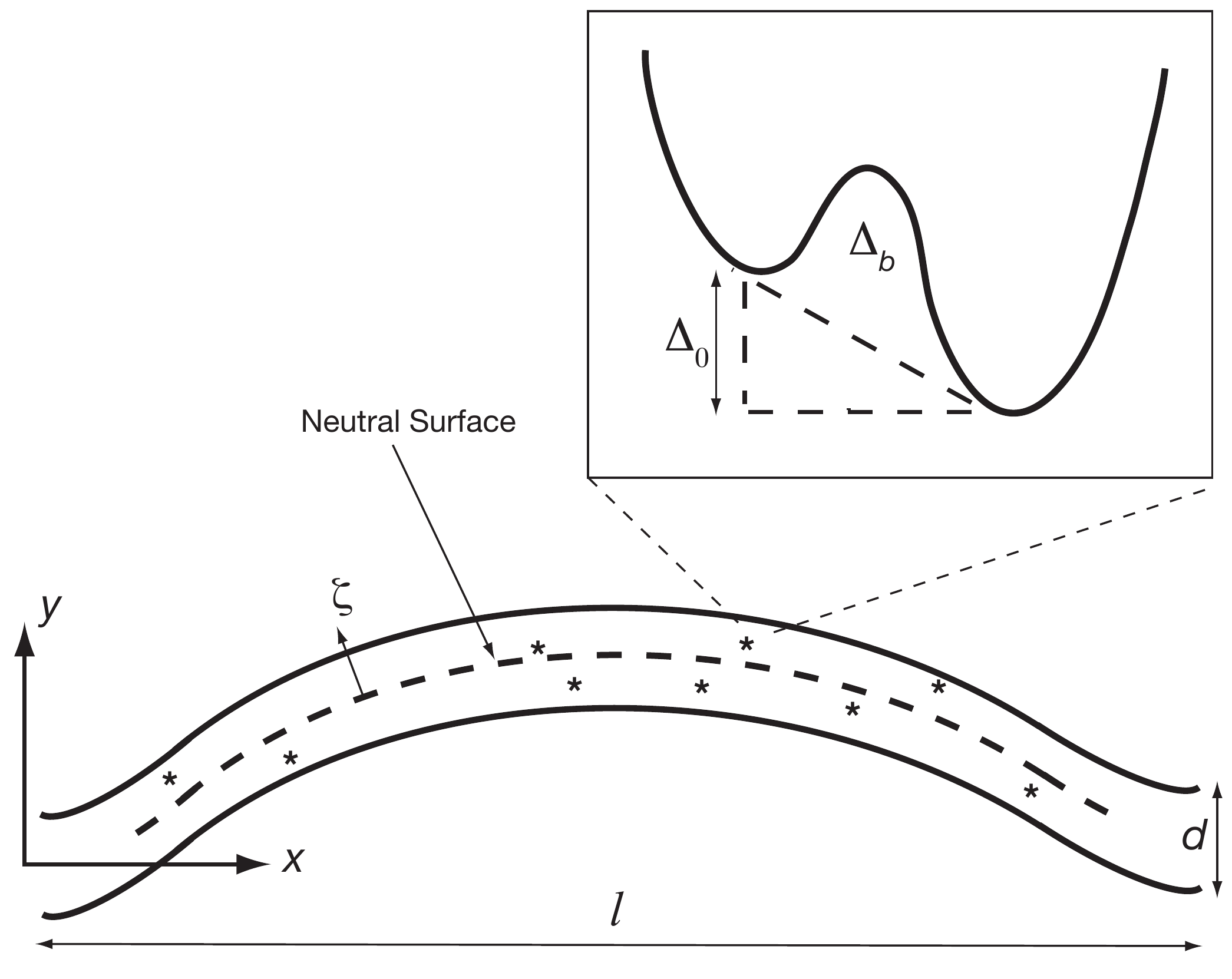}
	\caption{Flexing doubly-clamped beam (exaggerated for clarity). Inset: TLS double well potential.}
	\label{beamTLS}
\end{figure}

In this section we will derive the Hamiltonian describing the dynamics of the lowest, fundamental flexural mode of a doubly-clamped beam mechanical resonator interacting with TLS's that are located randomly throughout the beam volume. Related analyses are given in Refs.~[\onlinecite{kuhn07,seoanez08}]. We shall assume a long, thin elastically isotropic beam with length $l$, width $w$, and thickness $d$ satisfying $l\gg w>d$, mass density $\rho$, and bulk modulus $K$ (Fig.~\ref{beamTLS}). 
The equation of motion for small transverse displacements $y(x,t)$, $0<x<l$, of the beam is\cite{lawrence02}
\begin{equation}
\rho\frac{\partial^2 y}{\partial t^2}+K b^2 \frac{\partial^4 y}{\partial x^4}=0,
\label{resonatoreq}
\end{equation}
where $b=d/\sqrt{12}$ is the bending moment and we assume zero applied longitudinal strain. The total energy of the beam is
\begin{equation}
E=\frac{1}{2}\rho w d \int_0^l dx (\dot{y})^2 +\frac{1}{2}Kwd b^2 \int_0^l dx ({y}'')^2.
\label{totalenergyeq}
\end{equation}
Solving Eq.~(\ref{resonatoreq}) with clamped boundary conditions $y(0)=y(l)=y'(0)=y'(l)=0$, we obtain for the lowest frequency (fundamental) eigenmode:
\begin{equation}
y(x,t)=Y(t) \frac{\phi(x)}{\phi(l/2)},
\label{fundamentalsolutioneq}
\end{equation}
where the normalised eigenfunction is
\begin{equation}
\phi(x)=A\left\{\left[\sin\left(\frac{\pi\alpha x}{l}\right)-\sinh\left(\frac{\pi\alpha x}{l}\right)\right]+\beta\left[\cos\left(\frac{\pi\alpha x}{l}\right)-\cosh\left(\frac{\pi\alpha x}{l}\right)\right]\right\},
\label{eigenfunctioneq}
\end{equation}
with $\alpha\approx 1.51$ obtained from the clamped boundary condition expression $\cos(\pi\alpha)\cosh(\pi\alpha)=1$ and $\beta=[\cos(\pi\alpha)-\cosh(\pi\alpha)]/[\sin(\pi\alpha)+\sinh(\pi\alpha)]\approx -1.02$. The  constant $A\approx 0.983$ is fixed by requiring that  $\phi(x)$ be normalized as follows:
\begin{equation}
l^{-1}\int_0^l dx \phi^2(x)=1.
\label{normalizationcondeq}
\end{equation} 
The time-dependent part of the solution (\ref{fundamentalsolutioneq}) is
$Y(t)={\mathrm{Re}}(Y_0 e^{i\omega t})$, where the fundamental mode frequency is
\begin{equation}
\omega=b \left(\frac{\pi\alpha}{l}\right)^2\sqrt{\frac{K}{\rho}}\approx 6.46 \frac{d}{l^2}\sqrt{\frac{K}{\rho}}.
\label{fundamentalfreqeq}
\end{equation}
The solution~(\ref{fundamentalsolutioneq}) is expressed such that $Y(t)$ gives the transverse displacement of the beam at its midpoint $x=l/2$. Substituting Eq.~(\ref{fundamentalsolutioneq}) into the total energy~(\ref{totalenergyeq}) and employing the normalization condition~(\ref{normalizationcondeq}), we obtain
\begin{equation}
E=\frac{1}{2}m\phi^{-2}(l/2) \dot{Y}^2 +\frac{1}{2}m \phi^{-2}(l/2)\omega^2 Y^2,
\label{totalenergy2}
\end{equation}
where $m=\rho w d l$ is the mass of the resonator. Thus, the fundamental mode dynamics is that of   a harmonic oscillator with effective mass
\begin{equation}
m_{\mathrm{eff}}=\frac{m}{\phi^2(l/2)}\approx \frac{m}{2.52},
\label{effectivemasseq}
\end{equation}
where, from Eq.~(\ref{eigenfunctioneq}),
 we have used $\phi(l/2)\approx 1.59$.

Quantizing the fundamental mode, we introduce mode raising and lowering operators $\hat{a}^+$, $\hat{a}$; $[\hat{a},\hat{a}^+]=1$, with
\begin{equation}
\hat{Y} =Y_{zp}(\hat{a}+\hat{a}^+),
\label{Yoperatoreq}
\end{equation}
where $Y_{zp}=\sqrt{\hbar/(2 m_{\mathrm{eff}}\omega)}$ is the quantum zeropoint displacement uncertainty. The free beam, fundamental mode Hamiltonian is then simply $\hat{H}_m=\hbar\omega (\hat{a}^+\hat{a}+1/2)$.

Moving now to the TLS Hamiltonian, we have:
\begin{equation}
\hat{H}_{\mathrm{TLS}}=\sum_{j=1}^N \left[ \frac{1}{2}\Delta_0^{(j)}\sigma_z^{(j)}+\frac{1}{2}\Delta_b^{(j)}\sigma_x^{(j)}\right],
\label{TLSHamiltonianeq}
\end{equation}
where $j=1, 2,...,N$ labels the TLS,  $\Delta^{(j)}_0$  is the asymmetry of the $j$th TLS's  potential well and $\Delta_b^{(j)}$ is its tunnel splitting that depends on the well barrier height and width (see Fig.~\ref{beamTLS}).  Mechanical resonator motion couples to a TLS largely through the strain dependence of the asymmetry energy: 
\begin{equation}
\Delta_0^{(j)}[\epsilon_{kl}]\approx \Delta_0^{(j)} +2\sum_{k,l=1}^3 \nu_{k l}\epsilon^{(j)}_{k l}, 
\label{strainenergyeq}
\end{equation}
where $\nu_{kl}$ is the deformation potential and $\epsilon_{kl}$ is the elastic strain tensor.
For small amplitude, transverse $y(x,t)$ flexural displacements of long, thin beams, the  nonvanishing strain tensor components for a defect located at $x$ and a distance $-d/2\leq\zeta\leq d/2$ normal to the neutral (i.e., strain-free) surface (see Fig.~\ref{beamTLS}) are $\epsilon_{xx}\approx -\zeta d^2 y/dx^2$ and $\epsilon_{yy}=\epsilon_{zz}\approx \sigma\zeta d^2 y/dx^2$, where $\sigma$ is Poisson's ratio.    From Eqs.~(\ref{fundamentalsolutioneq}) and (\ref{Yoperatoreq}), the transverse displacement field operator is 
\begin{equation}
\hat{y}(x)=\frac{\phi(x)}{\phi(l/2)}Y_{zp}(\hat{a}+\hat{a}^+).
\label{yfieldoperator}
\end{equation}
Subsituting (\ref{yfieldoperator}) into  (\ref{strainenergyeq}), we obtain for the mechanical resonator-TLS defect interaction Hamiltonian
\begin{equation}
\hat{H}_{\mathrm{int}}=\sum_{j=1}^{N}\lambda^{(j)} (\hat{a}+\hat{a}^+)\sigma_z^{(j)},
\label{interactionhamiltonianeq}
\end{equation}
where the resonator-TLS strain coupling strength for defect $j$ located at $(x_j,\zeta_j)$ in the beam takes the form
\begin{equation}
\lambda^{(j)}=-Y_{zp}\nu (1-2\sigma)  \zeta_j\frac{\phi''(x_j)}{\phi(l/2)}
\label{couplingstrengtheq}
\end{equation}
and we have assumed for simplicity an isotropic deformation potential coupling $\nu$.
Finally, writing out the full resonator-TLS system Hamiltonian, we have
\begin{equation}
\hat{H}_S=\hbar\omega (\hat{a}^+\hat{a}+1/2) +\sum_{j=1}^N \left[ \frac{1}{2}\Delta_0^{(j)}\sigma_z^{(j)}+\frac{1}{2}\Delta_b^{(j)}\sigma_x^{(j)}+\lambda^{(j)} (\hat{a}+\hat{a}^+)\sigma_z^{(j)}\right].
\label{fullhamiltonianeq}
\end{equation}

The strength of the coupling $\lambda^{(j)}$ depends on the location of the TLS defect. In particular, the coupling is strongest for a defect on the surface at the beam ends, i.e., $\zeta_j=\pm d/2$ and $x_j=0, l$. In order to gain a sense of the expected magnitudes of the coupling, it is convenient to express the various beam material constants and dimensions in natural units. We obtain for the dimensionless coupling strength:
\begin{eqnarray}
\frac{\lambda^{(j)}}{\hbar\omega}&\approx -10^{-3}&\left(\frac{\zeta_j}{d}\right) \left(\frac{\phi''(x_j)}{\phi''(0)}\right)\left(\frac{\rho}{10^3{\mathrm{ kg m}}^{-3}}\right)^{1/4} \left(\frac{K}{10^{11}{\mathrm{ Nm}}^{-2}}\right)^{-3/4} \left(\frac{\nu(1-2\sigma)}{\mathrm{eV}}\right)\cr
&&\times \left(\frac{l}{\mu\mathrm{m}}\right)^{1/2}
 \left(\frac{w}{\mu\mathrm{m}}\right)^{-1/2} \left(\frac{d}{\mu\mathrm{m}}\right)^{-1},
\label{dimensionlesscouplingeq}
\end{eqnarray}
while the fundamental flexural mode frequency $f=\omega/(2\pi)$ expressed in natural units is:
\begin{equation}
\left(\frac{f}{\mathrm{GHz}}\right)\approx 1.3\left(\frac{\rho}{10^3{\mathrm{ kg m}}^{-3}}\right)^{-1/2} \left(\frac{K}{10^{11}{\mathrm{ Nm}}^{-2}}\right)^{1/2} \left(\frac{l}{\mu\mathrm{m}}\right)^{-2} \left(\frac{d}{\mu\mathrm{m}}\right).
\label{dimensionlessfrequencyeq}
\end{equation}

\section{\label{sec:open}Open system master equation}
In this section we derive the master equation for the coupled resonator-TLS system, taking into account the environment of the system. In an actual beam mechanical resonator, the fundamental flexural mode will couple not only to the TLS defects, but also to the other, higher frequency resonator modes via anharmonic interaction terms. The fundamental mode will also couple to bulk, substrate modes at the beam supports. Furthermore, the TLS's will couple to the higher frequency resonator modes through the strain dependence of the TLS's asymmetry energies. The latter will not only cause damping of the TLS's,\cite{seoanez08} but will also induce interactions between the TLS's.\cite{black77,burin98,anghel08} However, our goal in the present investigation is not to accurately model the respective environments of the fundamental mode and TLS's, but rather as a first step to consider the simplest possible idealized model environments in order to gain an idea of the quantum dissipation and decoherence dynamics of the mechanical resonator interacting with damped TLS's.

As idealised model environments, we consider baths of non-interacting harmonic oscillators. Hamiltonian~(\ref{fullhamiltonianeq}) is then augmented by the environment Hamiltonian and coupling term:
\begin{eqnarray}
{H}_{\mathrm{env}}+H_{S-{\mathrm{env}}}&=&\sum_n \hbar\omega_n {b}^+_n {b}_n + \sum_{j=1}^N \sum_n \hbar\omega_n {c}^{(j)+}_n {c}^{(j)}_n +Y\sum_n\kappa_n ({b}^+_n +{b}_n)\cr
&&+ \sum_{j=1}^N {\sigma}^{(j)}_z \sum_n \tilde{\kappa}^{(j)}_n ({c}^{(j)+}_n + {c}^{(j)}_n),\label{envhamiltonianeq}
\end{eqnarray}
where recall $Y=Y_{zp} (a +a^+)$ and for notational convenience we have dropped the hats on the operators and have also neglected the bath zeropoint contributions. In our model each TLS is assumed to couple with strength $\tilde{\kappa}^{(j)}_n$  to independent, noninteracting oscillator baths, characterized by environment mode operators $c_n^{(j)}, j=1,2,...,N$. The environment mode operators $b_n$ couple directly to the resonator with strength $\kappa_n$,  collectively modeling all  energy loss mechanisms other than those involving the TLS's, such as  clamping loss and anharmonic processes.  
Combining Hamiltonians (\ref{fullhamiltonianeq}) and (\ref{envhamiltonianeq}), we have the total system-environment Hamiltonian $H=H_S+H_{\mathrm{env}}+H_{S-{\mathrm{env}}}$.

In Appendix~\ref{sec:masterequation} we apply the self-consistent Born approximation together with a Markov approximation to obtain the following  master equation describing the dissipative dynamics of the coupled resonator-TLS system: 
\begin{eqnarray}
\dot{\rho}_S (t)&=&-\frac{i}{\hbar}[H_S,\rho_S(t)]-\frac{i\gamma}{2\hbar} [Y,\{ P_Y,\rho_S(t)\}]-\frac{m\omega\gamma}{2\hbar}\coth\left(\frac{\hbar\omega}{2k_B T}\right)[Y,[Y,\rho_S(t)]]\cr
&&-\sum_{j=1}^N\frac{1}{4T_1^{(j)}}\left(\frac{E^{(j)}}{\Delta^{(j)}_b}\right)^2[\sigma^{(j)}_z,[\sigma^{(j)}_z,\rho_S(t)]]
\cr&&-\sum_{j=1}^N\frac{i}{4T_1^{(j)}}\left(\frac{E^{(j)}}{\Delta^{(j)}_b}\right)\tanh\left(\frac{E^{(j)}}{2 k_B T}\right)[\sigma^{(j)}_z,\{\sigma^{(j)}_y,\rho_S(t)\}],
\label{quantumBMMastereq}
\end{eqnarray}
where $\rho_S(t)$ is the resonator-TLS system density matrix, $P_Y$ is the resonator momentum, $\{\cdot,\cdot\}$ denotes the anticommutator and $E^{(j)}=\sqrt{(\Delta_0^{(j)})^2+(\Delta_b^{(j)})^2}$ is the $j$th TLS energy level separation. The parameter $\gamma$ gives the energy damping rate of the resonator in the absence of the TLS, while $T_1^{(j)}$ gives the $j$th TLS relaxation time from its excited energy eigenstate in the absence of the resonator. 

One potential advantage of the damping time/rate parametrization  used in Eq.~(\ref{quantumBMMastereq}) is that it does not in fact depend explicitly on the microscopic nature of the environment and how it couples to the system. As long as the Markov approximation can be made and the fluctuation-dissipation theorem holds, then the damping and diffusion terms are uniquely related so that Eq.~(\ref{quantumBMMastereq}) can be assumed to apply for other system-environment interactions as well. This then allows environment model-independent predictions for the open system dynamics  provided that they are expressed in terms of the damping parameters $\gamma$ and $T^{(j)}_1$, as opposed to predictions concerning the explicit temperature dependence,\cite{seoanez08} which depend on the nature of the TLS environment.

\section{\label{sec:damping} Damping} 
In this and the following section, we present the results of  numerically solving  the master equation~(\ref{quantumBMMastereq}) using the Quantum Optics Toolbox.\cite{tan} Most of the  results are for  the mechanical resonator mode subsystem only with the TLS sector of state space traced over;  we assume that  it is the resonator mode which  is directly probed in experiment. 

We begin this section with a focus on the damping of the mechanical resonator  coupled to a single TLS.  Both the resonator and the TLS are coupled to  independent ohmic oscillator baths.  We assume that the system is initially in a product state, $\rho_S(0)=\rho_{\mathrm{res}}(0)\otimes\rho_{\mathrm{TLS}}(0) $, where the TLS is initially in a thermal state $\rho_{\mathrm{TLS}}(0)=\exp [-H_{\mathrm{TLS}} /k_BT] /Z $ with $H_{\mathrm{TLS}}$ defined in Eq.~(\ref{TLSHamiltonianeq}) and $Z=\mathrm{Tr}( \exp [-H_{\mathrm{TLS}} /k_BT] )$.  Similarly, the initial resonator state is a thermal state with $H_m=\hbar\omega (a^{\dagger} a+1/2)$ that has been displaced using the operator $d = \exp [x_0(a^{\dagger}-a)/2 ]$, where  $x_0$ is the initial displacement of the thermal state from equilibrium in units of the quantum zero-point displacement uncertainty $Y_{zp}$ defined in Eq.~(\ref{Yoperatoreq}); from now on we use $x$    to denote the mechanical resonator fundamental mode displacement in units of  the zero-point uncertainty. 

We use Eqs.~(\ref{dimensionlesscouplingeq}) and (\ref{dimensionlessfrequencyeq}) to determine the resonator-TLS coupling constant $\lambda$ and the fundamental flexural mode frequency $f$.  For $l=10 ~\mu \mathrm{m},\;w=1~ \mu \mathrm{m},\;d=0.1~ \mu \mathrm{m}$, we find $\lambda/\hbar\omega\approx 10^{-2}$ and $f\approx100~\mathrm{MHz}$, or $\omega\approx2\pi\times100~\textrm{MHz}$. Where convenient, we shall use dimensionless time units, $t\rightarrow\omega t$, with  $T_1$ and $\gamma$  expressed as $\omega T_1$ and  $\gamma/\omega$, respectively,  and $\lambda$, $\Delta_i$, and temperature $T$  expressed as $\lambda/\hbar\omega$, $\Delta_i/\hbar\omega$, and $k_BT/\hbar\omega$, respectively.

\begin{figure}[htbp]
	\centering
		\includegraphics[height=3in]{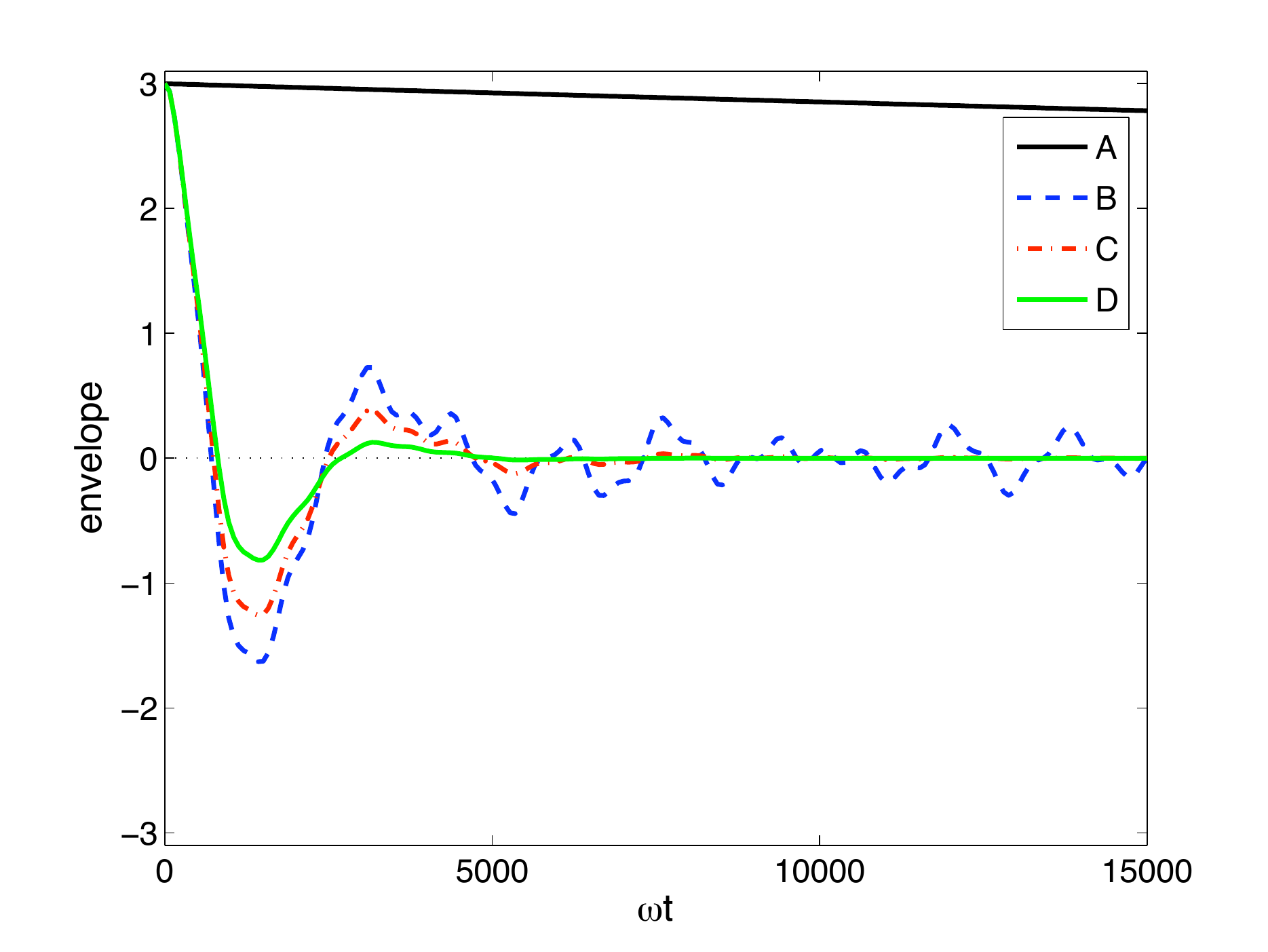}
	\caption{Envelope of the motion of the resonator coupled to an on resonance TLS for $T_1\approx6.5\times10^4$ (B), $T_1\approx1.6\times10^4$ (C), and $T_1\approx6.5\times10^2$ (D).  The solid black curve (A) shows the damping of the resonator with no TLS coupling.  For all curves $\lambda=10^{-2},\;\Delta_0=0,\;\Delta_b=1,\; \gamma=10^{-5},\;T=2.02.$}
	\label{damping:figure1}
\end{figure}

We begin with a symmetric TLS ($\Delta_0=0$) that is on resonance with the mechanical oscillator ($\Delta_b=1$).  In Fig.~\ref{damping:figure1}, we plot the envelope of the resonator's  ensamble averaged position versus time, corresponding to the so-called interaction picture or rotating frame, where the resonator and TLS's rapid free evolution are factored out. We give the resonator  an initial displacement $x_0=3$ and assume a range of  experimentally realistic values for the TLS relaxation time that are relevant at mK temperatures, $1~\mu\textrm{sec}\lesssim T_1\lesssim 100~\mu \textrm{sec} $,\cite{graebner79} and also assume a non-TLS resonator energy damping rate $\gamma=10^{-5}$.  Comparing the envelope of the resonator's motion at $T=10\textrm{ mK}$ ($=2.02$ in dimensionless units) for $T_1=100~\mu \textrm{sec}$ ($=6.5\times10^4$ in dimensionless units) (curve B) to the envelope of the resonator in the absence of the TLS (curve A), we see that the TLS and its bath cause significant amplitude damping of the resonator, even for large values of $T_1$.  Fig.~\ref{damping:figure1} also shows an increase in the damping of the resonator for decreasing values of $T_1$; for shorter TLS damping times, energy exchanged between the resonator and the TLS is more quickly dissipated through the TLS bath.  

\begin{figure}[htbp]
	\centering
		\includegraphics[height=3in]{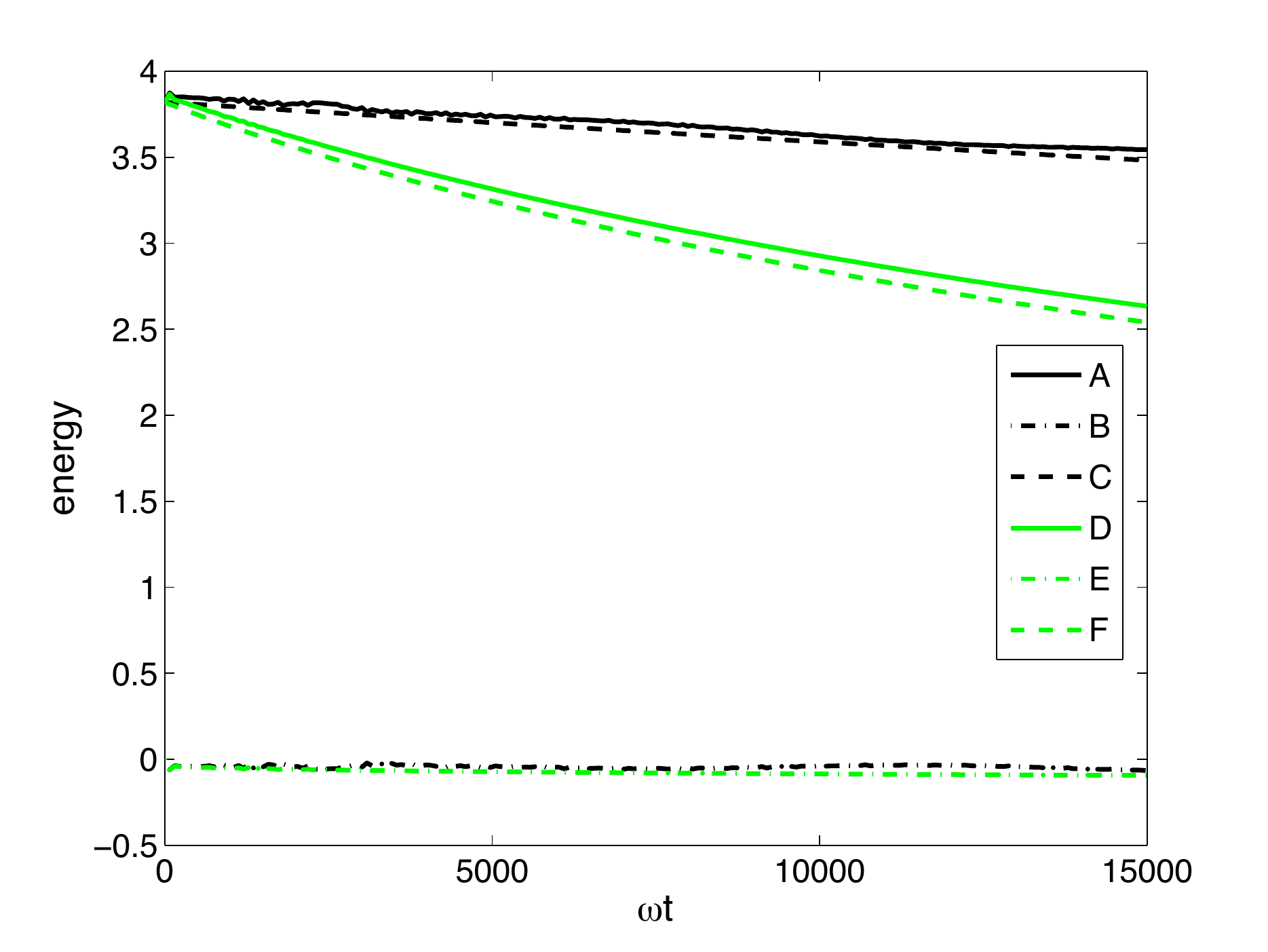}
	\caption{Dependence of resonator and TLS average energy vs. time for the same parameters as those used in Fig.~\ref{damping:figure1}, with $T_1\approx 6.5 \times 10^4$ (black) and $T_1\approx6.5\times10^2$ (green). (A, D) Resonator energy; (B, E) TLS energy; (C, F) Resonator + TLS energy. }
	\label{energydamping:fig}
\end{figure}

Note also from Fig.~\ref{damping:figure1} that the resonator's average position damps out and then revives on time scales much less than the TLS relaxation time, $T_1$, appearing to indicate a complete transfer of  energy from the mechanical resonator to the  TLS. However, the on-resonance TLS can only absorb a maximum of one quantum (phonon) of vibrational energy, while the  energy stored in the resonator for the considered initial displacement $x_0=3$ corresponds to more than one phonon on average. This apparent paradox is resolved by noting that the quantum resonator's average energy depends on the average of the position squared, not the square of the average; not all of the energy is transferred to the TLS when the average position completely damps out. This can be seen in Fig.~\ref{energydamping:fig}, where the individual resonator and TLS energies, as well as the total TLS+resonator energy, are plotted over time. 

\begin{figure}[htbp]
	\centering
		\includegraphics[height=3in]{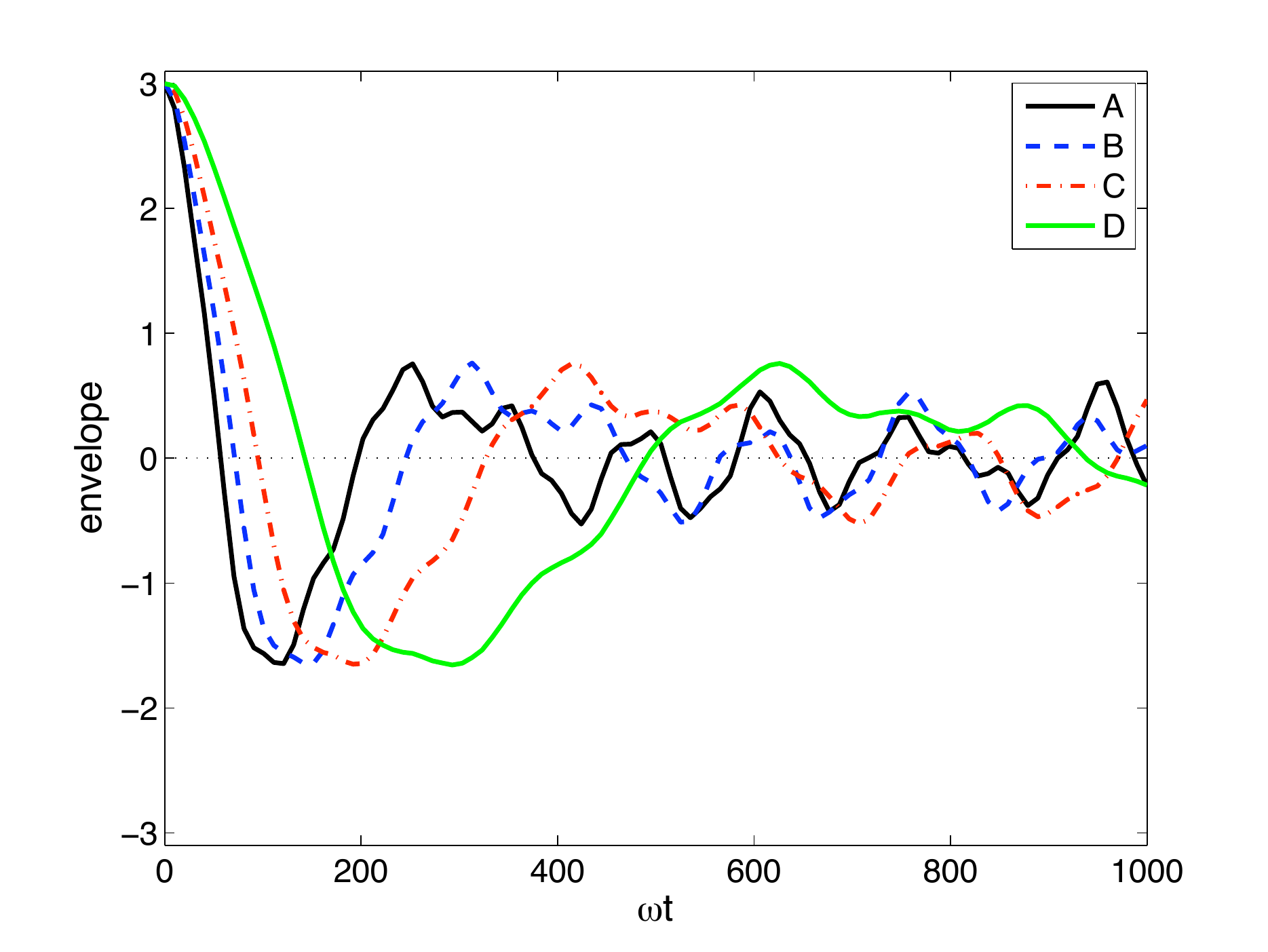}
	\caption{Dependence of the  resonator ensemble-averaged amplitude vs time for different resonator-TLS coupling constants $\lambda$.  The parameters are the same as those used in Fig.~\ref{damping:figure1}, with $T_1\approx6.5\times10^4$, $\lambda=0.125$ (A), $\lambda=0.1$ (B), $\lambda$=0.075 (C), and $\lambda=0.05$ (D).}
	\label{damping:figure4}
\end{figure}

Energy is able to be transfered between the oscillator and the TLS because of the long TLS $T_1$ times.  If $T_1$ is shorter then the energy transferred to the TLS dissipates to its bath more rapidly, with the result that less energy returns to the oscillator.  The effect of a shorter $T_1$ on the oscillator and oscillator+TLS energy is shown in Fig.~\ref{energydamping:fig}.  Fig.~\ref{damping:figure4} demonstrates the dependence of the resonator dynamics on the oscillator-TLS coupling constant $\lambda$.  For weaker couplings the damping out and  revival of the oscillator's motion occurs at later times.

\begin{figure}[htbp]
	\centering
		\includegraphics[height=2.3in]{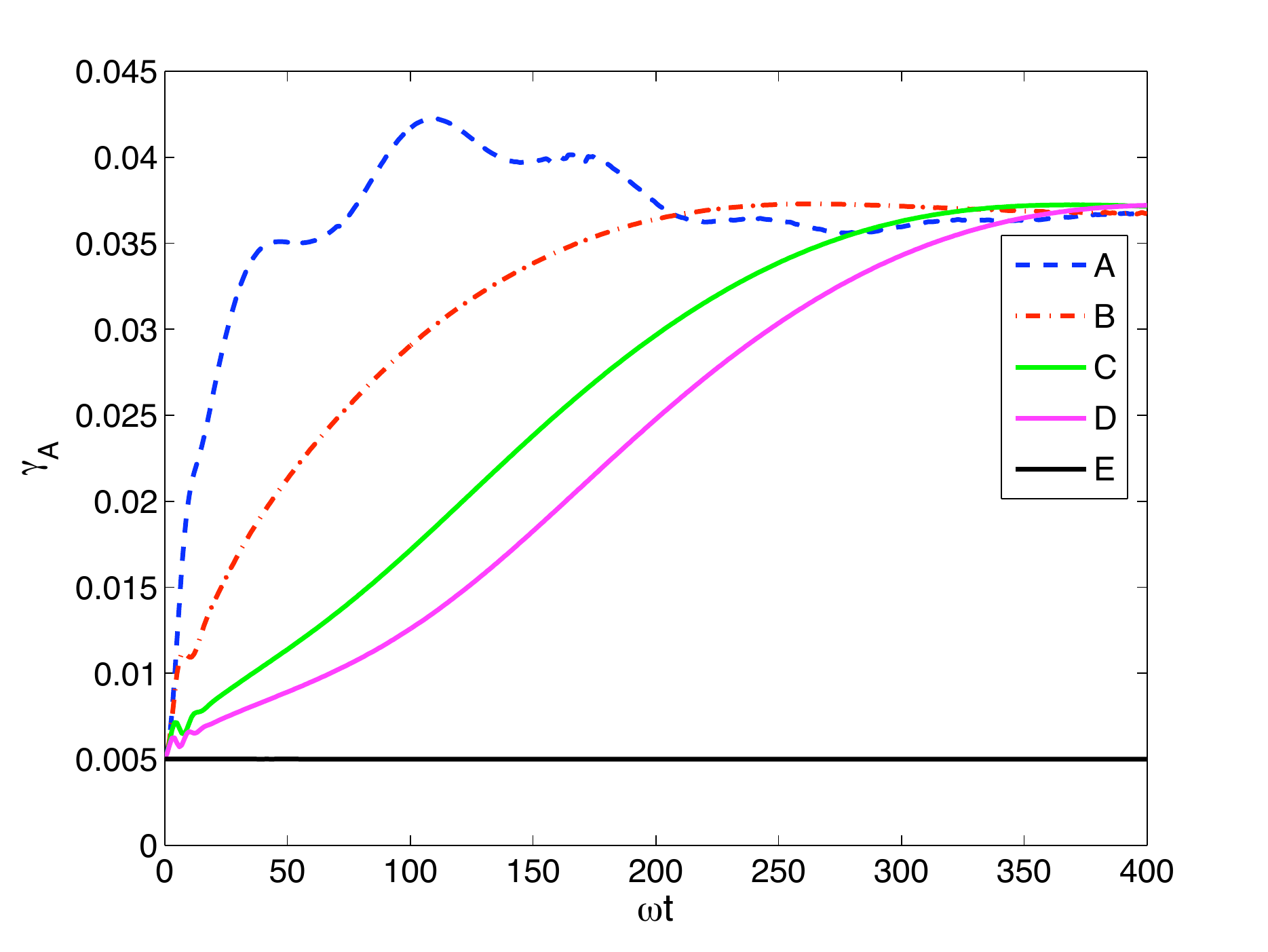}
		\includegraphics[height=2.3in]{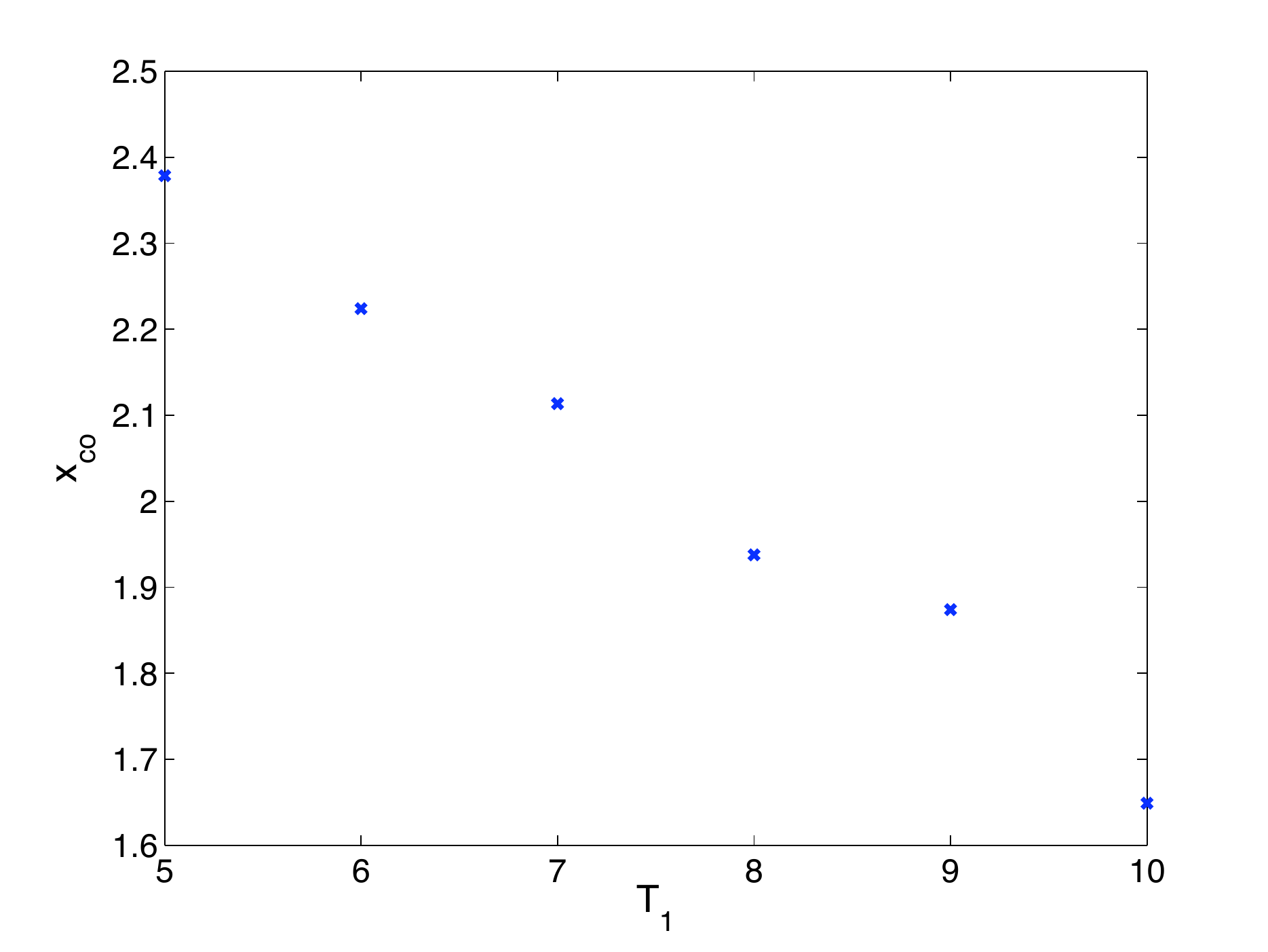}
	\caption{Left: Amplitude decay rate of resonator coupled to TLS with $T_1=10$ for $x_0=3$ (A), $x_0=5$ (B), $x_0=8$ (C), $x_0=10$ (D) and without TLS (E) as a function of time.  Right: Dependence of crossover amplitude $x_{\mathrm{CO}}$ on $T_1$ for $x_0=8$.  For all curves $\lambda=0.1,\;\Delta_0=0,\;\Delta_b=1,\;\gamma=0.01,\;\mathrm{and}\;T=1.$}
	\label{damping:figure2}
\end{figure}

Because a TLS absorbs only a finite amount of energy $E=\sqrt{\Delta_0^2 +\Delta_b^2}$, it can become saturated and hence affect the resonator damping rate.  In Fig.~\ref{damping:figure2} we plot the resonator amplitude decay rate $\gamma_A$ versus time for a range of initial displacements $x_0$, where $\gamma_A$ is defined as follows:
\begin{equation}
\gamma_A=-\frac{\partial}{\partial (\omega t)}\ln\left(\langle x(\omega t)\rangle\right).
\end{equation}
In order to display trends in damping more clearly we have increased the values for $\gamma$, $\lambda$ and $T_1^{-1}$, shortening the time over which the damping takes place.  We see in Fig.~\ref{damping:figure2} that as $x_0$ increases, the initial decay rate of the resonator decreases.  
For a large initial displacement, $x_0=10$ (D), the initial decay rate approaches that of the resonator in the absence of the TLS (E), suggesting near total saturation of the TLS.  For later times, however, when the resonator amplitude has decayed to a value near the zero point displacement, the decay rates for all four initial displacements approach a common,  constant value.  For $x_0=5$ this value is reached at $\omega t \approx 250$, while for $x_0=10$ this value is reached at $\omega t \approx 375$.  We know that the oscillator bath's contribution to damping is constant at all times (Fig.~\ref{damping:figure2}[E]), whereas the resonator coupled to the TLS shows amplitude-dependent damping initially, and amplitude-independent damping at later times (Fig.~\ref{damping:figure2}[A-D]).  These two distinct behaviors indicate that the TLS is indeed saturated at higher resonator amplitudes, while for lower amplitudes the TLS is unsaturated and its contribution to damping is therefore uniform.  On the right-hand side of Fig.~\ref{damping:figure2} we investigate the $T_1$ dependence of the amplitude at which the crossover from amplitude-dependent to amplitude-independent damping occurs. The crossover amplitude $x_{\mathrm{CO}}$ is plotted as a function of $T_1$, showing a nearly linear relationship.  

\begin{figure}[htbp]
	\centering
		\includegraphics[height=2.3in]{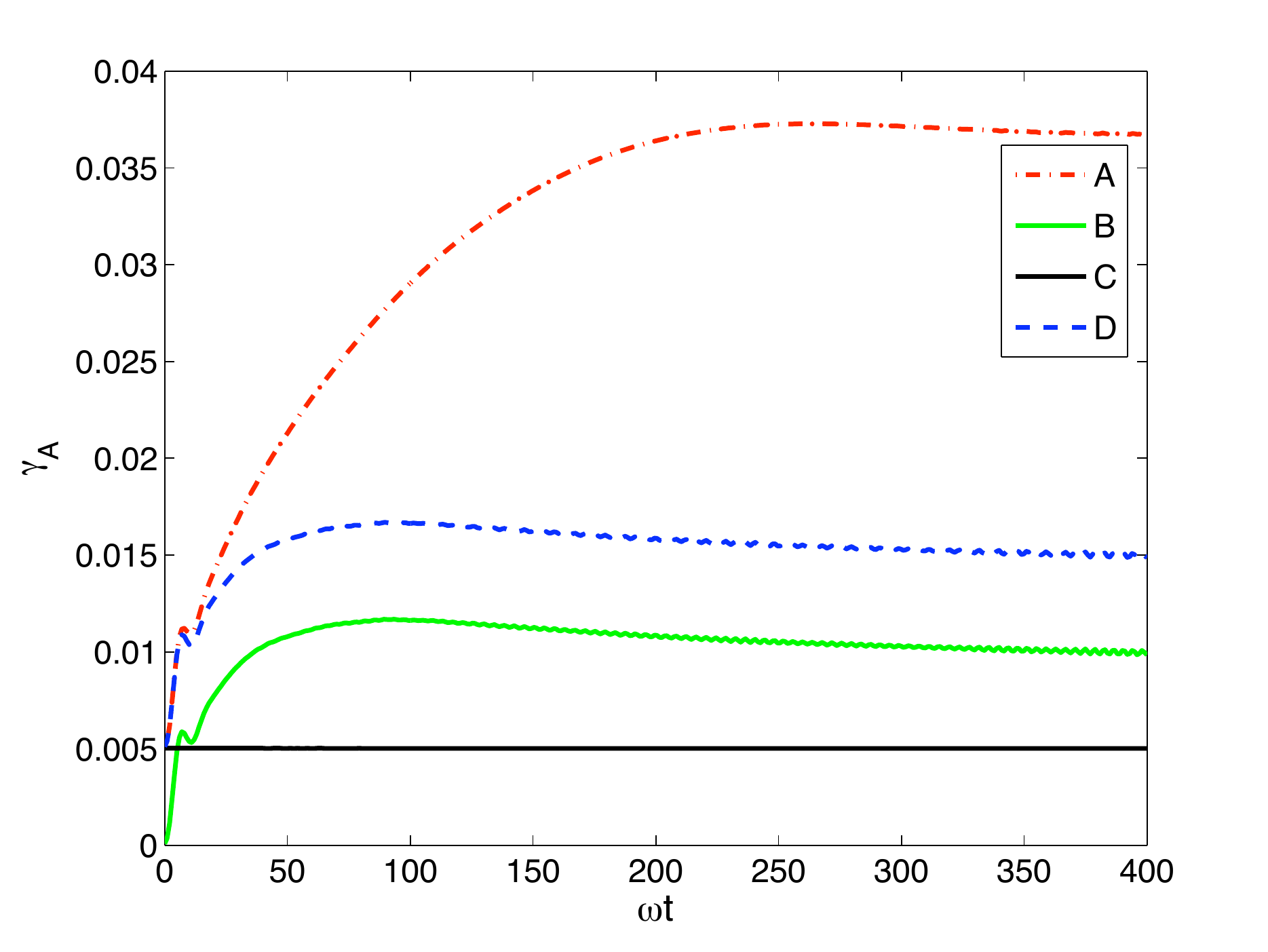}
	\caption{Decay rate of resonator for $x_0=5$ with TLS and oscillator bath (A), with TLS only (B), with oscillator bath only (C), and sum of curves B and C (D).   For all curves $\lambda=0.1,\;\Delta_0=0,\;\Delta_b=1,\;\gamma=0.01,\;T=1,\;\mathrm{and}\; T_1=10.$}
	\label{damping:cantadd}
\end{figure}

Fig.~\ref{damping:cantadd} shows the damping of a resonator with initial displacement $x_0=5$  coupled to both an oscillator bath and a TLS (A), and  coupled to the TLS (B) and oscillator bath (C) individually.  Curve D is the sum of curves B and C. The substantial difference between curves A and D demonstrates that one cannot simply add the individual damping rates to obtain the net TLS+oscillator bath damping rate when both these sources are present. 

\begin{figure}[htbp]
	\centering
		\includegraphics[height=3in]{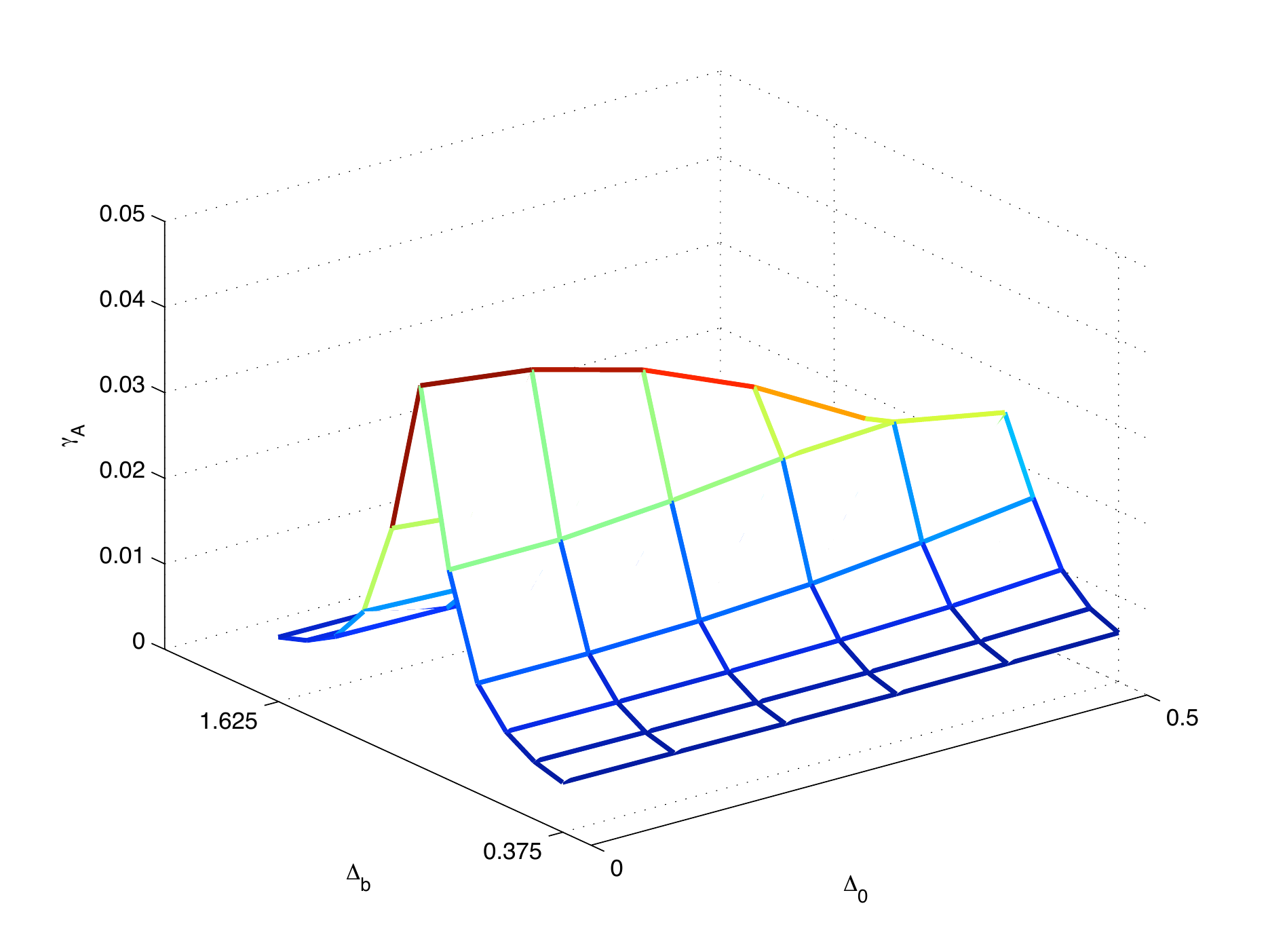}
	\caption{Resonator damping rate for various values of $\Delta_0$ and $\Delta_b$. The parameter values  $\lambda$, $\gamma$, $T$, and $T_1$  are the same as those used in Fig.~\ref{damping:figure2}.}
	\label{damping:figure3}
\end{figure}

So far we have considered just the special case of a symmetric ($\Delta_0=0$) and resonant ($\Delta_b=1$) TLS. In an actual mechanical resonator there will be a distribution of TLS's, not necessarily symmetric or on resonance [see Eq.~(\ref{distributioneq})].  In Fig.~\ref{damping:figure3}, we  show the resonator's amplitude damping rate $\gamma_A$ for a range of TLS $\Delta_0$ and $\Delta_b$ values.  In the absence of a TLS, $\gamma_A$ is equal to half the energy damping rate, $\gamma$, due to the oscillator bath.  Fig.~\ref{damping:figure3} shows that $\gamma_A$ is greatest for $\Delta_0=0$ and $\Delta_b$=1, and sharply decreases to $\gamma/2$ as $\Delta_b$ is moved off resonance.  The resonator's damping rate decreases more gradually as the TLS is made more asymmetric ($\Delta_0\neq0$).  The damping rates in this figure were extracted from an oscillator with small initial displacements in the range of constant values shown in Fig.~\ref{damping:figure2}, and therefore showed no time or amplitude dependence.

\begin{figure}[htbp]
	\centering
		\includegraphics[height=3in]{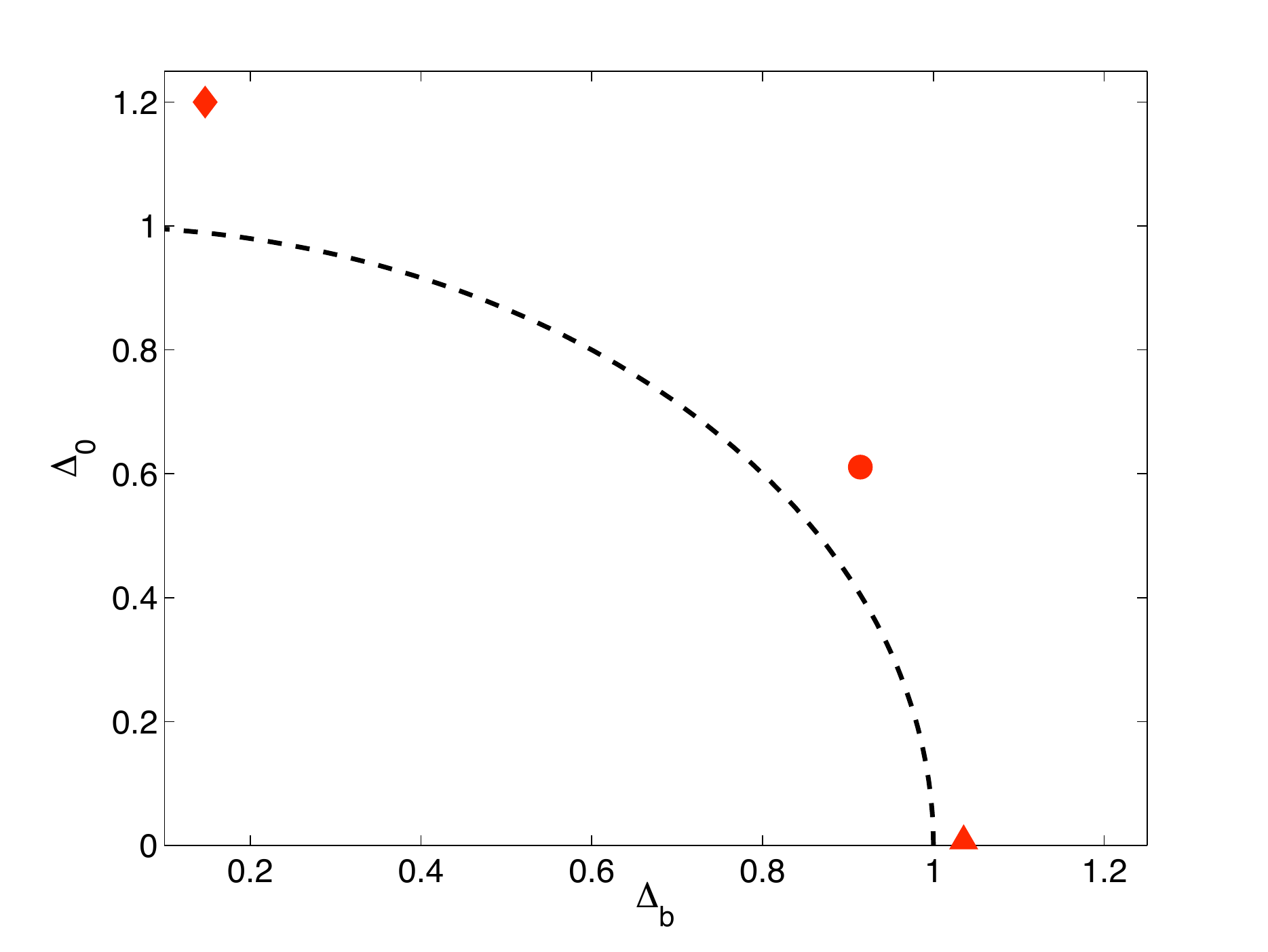}
	\caption{Considered distribution of TLS's. The energies of the TLS's  are $\Delta_0=1.2002$ and $\Delta_b=0.1475$ (diamond), $\Delta_0=0.6108$ and $\Delta_b=.9145$ (circle), $\Delta_0=0.0070$ and $\Delta_b=1.0354$ (triangle).  The dashed line denotes the  TLS energy on resonance with the mechanical oscillator.}
	\label{quadrant}
\end{figure}

\begin{figure}[htbp]
	\centering
		\includegraphics[height=3in]{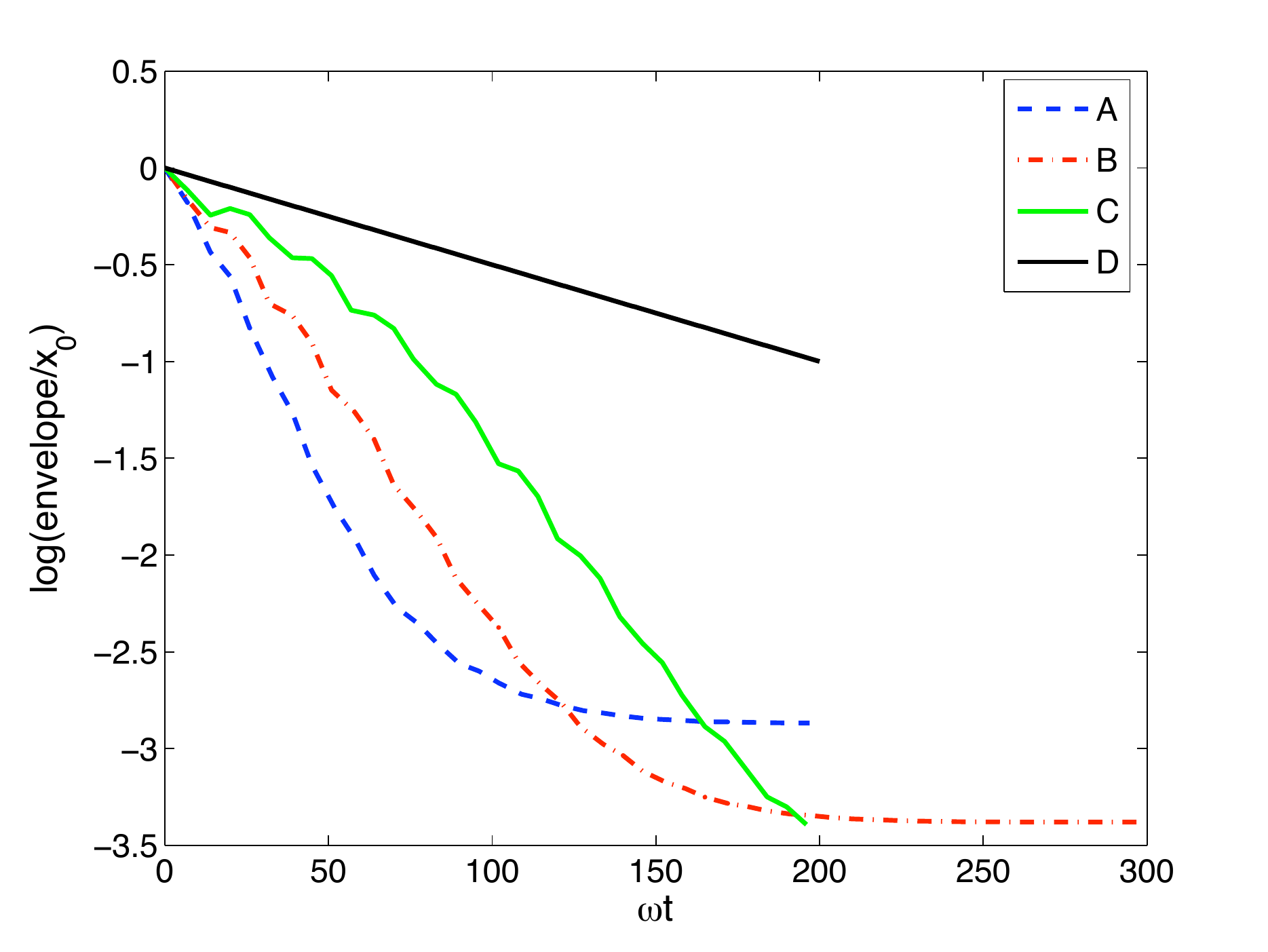}
	\caption{Damping of the oscillator coupled to three TLS's with initial displacements  $x_0=3$ (A), $x_0=5$ (B), and $x_0=8$ (C).  For all curves $\lambda=0.1,\;\gamma=0.01,\;T=1,\;\mathrm{and}\;T_1=10.$  (D) shows the damping of the oscillator in the absence of the TLS's. }
	\label{threeTLS:figure1}
\end{figure}

We now consider the damping of the resonator coupled to three TLS's, which have energies close to the resonator energy.
The TLS's  $\Delta_i$ values are indicated by the diamond,  circle, and triangle symbols in Fig.~\ref{quadrant}. The TLS energies were selected randomly from within the range $0.75\leq E\leq 1.25$ using the distribution~(\ref{distributioneq}).
As is the case for a single TLS, we expect the damping of the resonator when coupled to multiple TLS's to show amplitude dependence for early times due to TLS saturation.  Fig.~\ref{threeTLS:figure1} shows the decay in the presence of the three TLS's and the oscillator bath for three different initial displacements $x_0$ (curves A, B, C). As expected, the resonator decays more quickly for smaller initial displacements, while for larger initial displacements the decay rate approaches that due to the oscillator bath only, indicating saturation. Unexpectedly, at later times there is a crossover to a much slower, constant decay rate that is even less than that of the resonator in the absence of the TLS's, i.e., due to the oscillator bath only. The unnormalized amplitudes of the envelope at which the crossover occurs are the same for each curve, independent of the initial $x_0$ displacement. We have also found similar crossover behavior in amplitude decay rates assuming other randomly selected distributions of three TLS's with energies close to the resonator energy.  The bumps in curves B and C are a result of numerical approximation and are not a product of the system's behavior.
 
\section{\label{sec:decoherence} Decoherence}

\begin{figure}[htbp]
	\centering
		\includegraphics[height=1.3in]{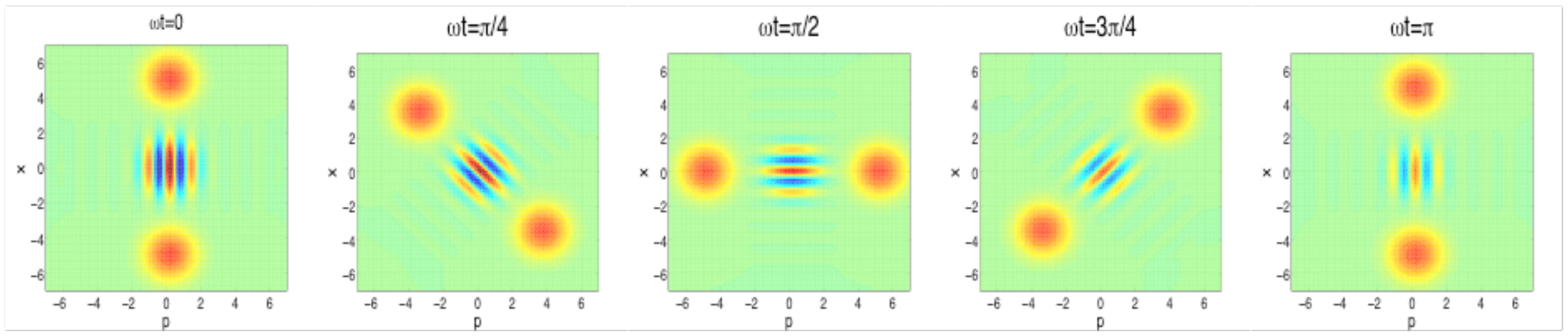}
	\caption{ Wigner function of the resonator density matrix in the absence of the TLS, where  $x_0=5,\;\gamma=0.01,\;T=1,\;\mathrm{and}\; T_1=10.$}
	\label{Wigner:figure1}
\end{figure}

\begin{figure}[htbp]
	\centering
		\includegraphics[height=1.3in]{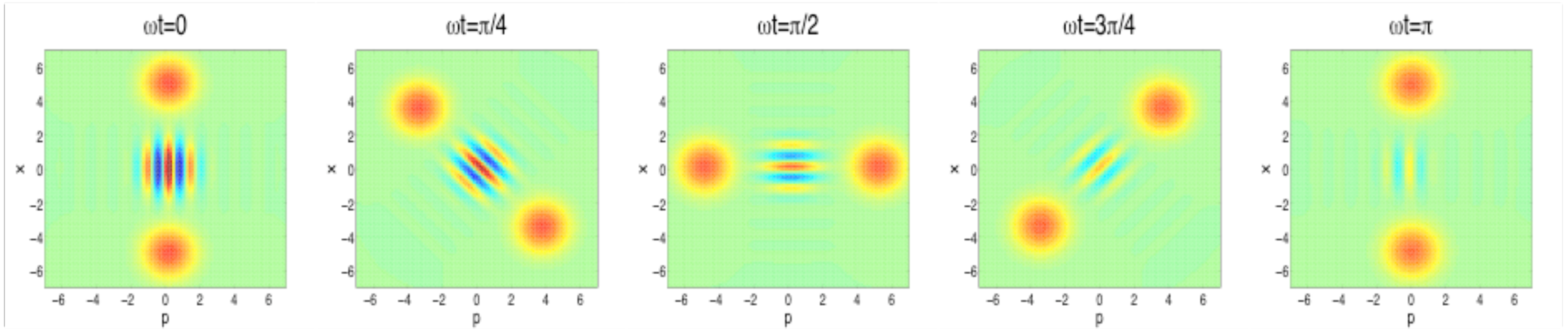}
	\caption{ Wigner function for the resonator coupled to the TLS, where  $x_0=5,\;\lambda=0.1,\;\Delta_0=0,\;\Delta_b=1,\;\gamma=0.01,\;T=1,\;\mathrm{and}\; T_1=10.$}
	\label{Wigner:figure2}
\end{figure}

\begin{figure}[htbp]
	\centering
		\includegraphics[height=1.3in]{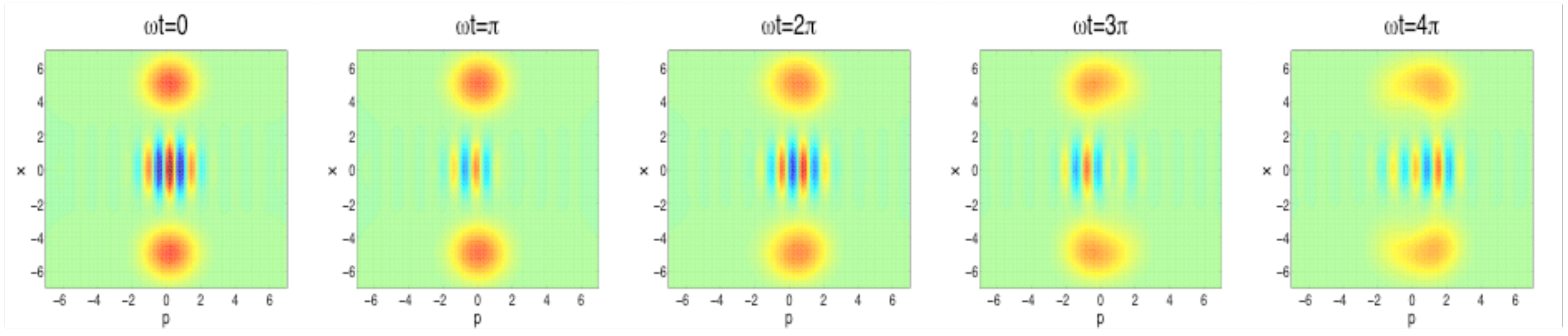}
	\caption{ Wigner function for the resonator coupled to the TLS, but without the resonator's independent ohmic oscillator bath, where  $x_0=5,\;\lambda=0.1,\;\Delta_0=0,\;\Delta_b=1,\;T=1,\;\mathrm{and}\;T_1=10.$  Note: time steps are larger than for Figs.~\ref{Wigner:figure1} and \ref{Wigner:figure2}.}
	\label{Wigner:figure3}
\end{figure}

In this section we investigate the contribution of the TLS's to the decoherence of a mechanical resonator Schr\"{o}dinger cat state.  The resonator's initial state consists of a superposition of two coherent states: $|\psi\rangle=N^{-1}\left(|+x_0\rangle+|-x_0\rangle\right)$, where $|\pm x_0\rangle=\exp(\pm x_0(\hat{a}^\dagger-\hat{a})/2)|0\rangle$ with initial displacement $\pm x_0$ and $N$ is a normalization factor.  The parameter choices are $\gamma=0.01$, $\lambda=0.1$ and $T_1=10$, the same as those used for showing the trends in damping in Sec.~\ref{sec:damping}.  In order to study the evolution of the decoherence of the cat state, we evaluate the Wigner function of the resonator density matrix $\rho$.  Fig.~\ref{Wigner:figure1} shows the Wigner function  in the absence of TLS's, i.e., with only the  ohmic oscillator bath, at five equal time step increments.  Within a single period of the oscillator's motion, the interference fringes between the two cat states decay substantially for the assumed parameter choices.  Fig.~\ref{Wigner:figure2} shows the evolution of the Wigner function for the same initial resonator state but with a single, on-resonance TLS coupled to the resonator in addition to the ohmic bath.  The interference fringes  decay more rapidly due to the presence of the TLS.  Fig.~\ref{Wigner:figure3} shows the Wigner function for the resonator coupled to a single TLS, but in the absence of the ohmic oscillator bath.  The decoherence time in this case is  longer than the  resonator's oscillation period.  

\begin{figure}[htbp]
	\centering
		\includegraphics[height=3in]{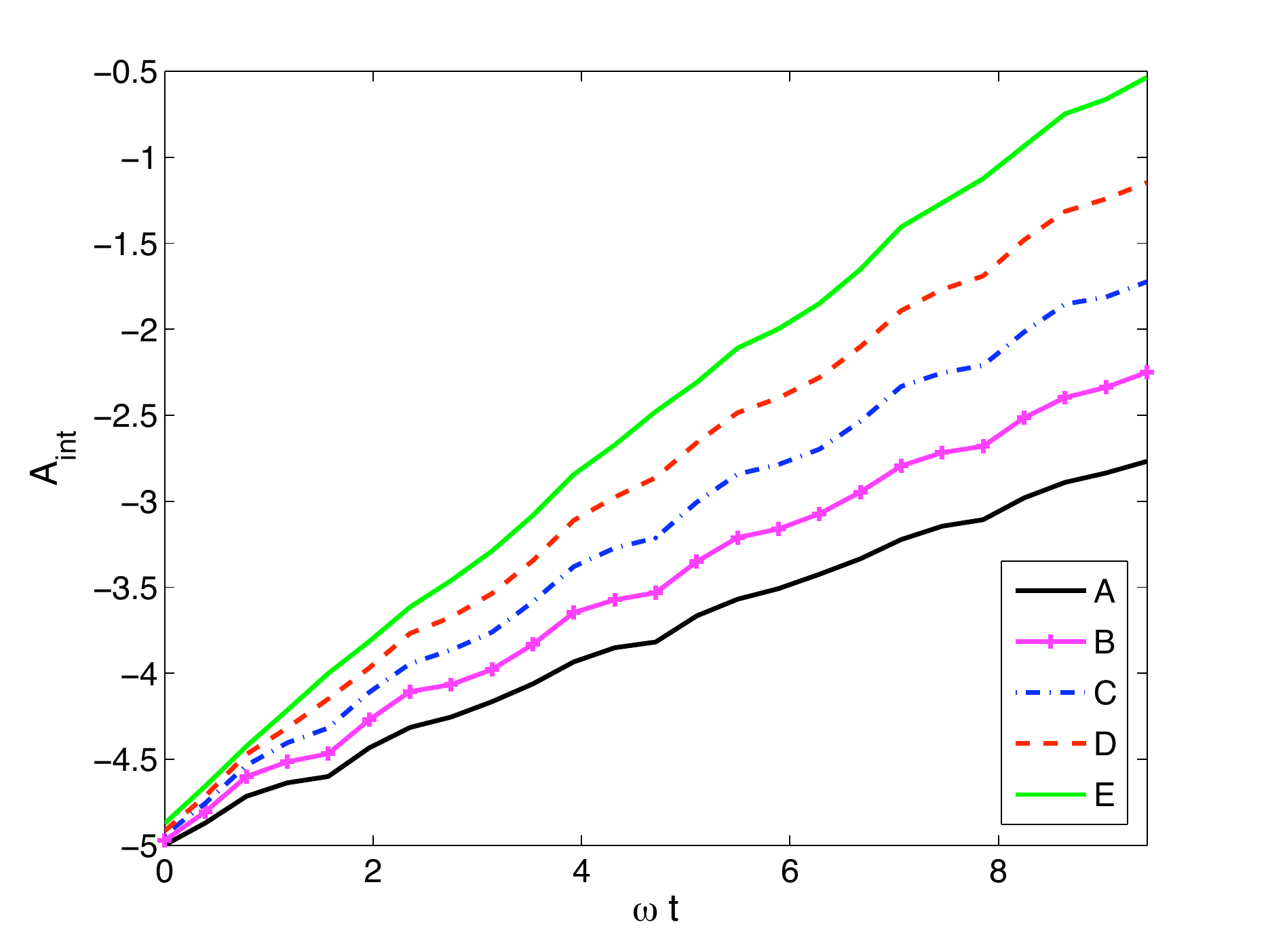}
	\caption{$A_{\mathrm{int}}$ vs. $\omega t$ for the resonator in the absence of TLS's.  Initial displacements are $x_0=5$~(A), $x_0=5.5$~(B), $x_0=6$~(C), $x_0=6.5$~(D), $x_0=7$~(E), with $\;\Delta_0=0,\;\Delta_b=1,\;\gamma=0.01,\;T=1,\;\mathrm{and}\;T_1=10.$}
	\label{A_intOsc}
\end{figure}
\begin{figure}[htbp]
	\centering
		\includegraphics[height=3in]{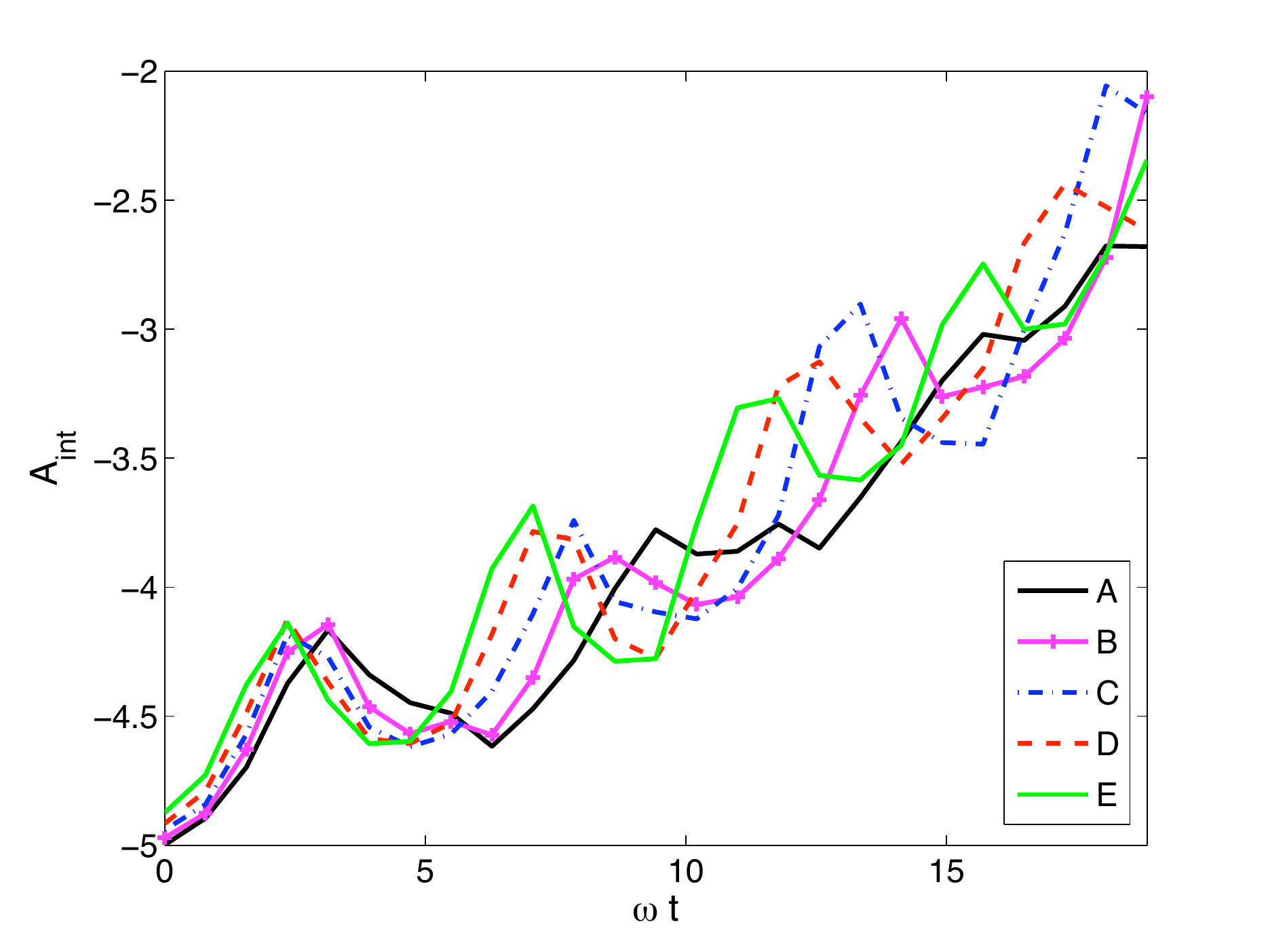}
	\caption{$A_{\mathrm{int}}$ vs. $\omega t$ for the resonator coupled to a single, on-resonance TLS, but not to the resonator's independent ohmic oscillator bath.  Initial displacements are $x_0=5$~(A), $x_0=5.5$~(B), $x_0=6$~(C), $x_0=6.5$~(D), $x_0=7$~(E), with  $\lambda=0.1$, $\;\Delta_0=0,\;\Delta_b=1,\;T=1,\;\mathrm{and}\;T_1=10.$}
	\label{A_intSpin}
\end{figure}
In order to obtain a more quantitative understanding of the TLS-induced decoherence, we take the average amplitude of the interference fringes in a small, disk-like region centered between the two peaks, and plot the negative log of this amplitude, which we denote  $A_{\mathrm{int}}$, as a function of time.  This quantity is essentially the same as that used in Ref.~[\onlinecite{paz92}], where they show that the decoherence rate  of a harmonic oscillator cat state coupled to an ohmic oscillator bath is proportional to  the slope of $A_{\mathrm{int}}$.   Fig.~\ref{A_intOsc} shows $A_{\mathrm{int}}$  as a function of time for the resonator in the absence of the TLS.   As predicted by theory, we can extract a decoherence rate from the constant increase of $A_{\mathrm{int}}(\omega t)$.  We find that the decoherence  rate, or the slope of $A_{\mathrm{int}}(\omega t)$, goes as the square of the initial displacement, as expected.\cite{zurek91}  Fig.~\ref{A_intSpin} shows $A_{\mathrm{int}}$ as a function of time for the resonator interacting with an on-resonance TLS, but without the resonator's independent ohmic oscillator bath.  The decoherence due to the TLS does not show the same dependence on the initial displacement as the oscillator bath-induced decoherence; in fact, there is  no apparent systematic dependence on initial displacement. In particular, increasing the  initial displacement does not result in an overall increase in the decoherence rate.

In the Wigner function plot  Fig.~\ref{Wigner:figure3} for the resonator coupled to the damped TLS only,  the interference fringes in the center region partially decay away and then return after a full mechanical period. These interference oscillations, which can also be seen in Fig.~\ref{A_intOsc}, are a consequence of the on-resonance TLS behaving like a position measuring device. As the resonator interacts with the TLS, the two distinct position states in the superposition cause the  TLS state to evolve in different ways, resulting in  an entangled resonator-TLS  state; the resonator partially decoheres. However, because the coupling between the TLS and its environment as parametrized by the $T_1$ time is relatively weak, subsequent evolution almost completely undoes the entanglement;  the resonator partially recoheres, restoring the interference between the two position states. 

\begin{figure}[htbp]
	\centering
		\includegraphics[height=3in]{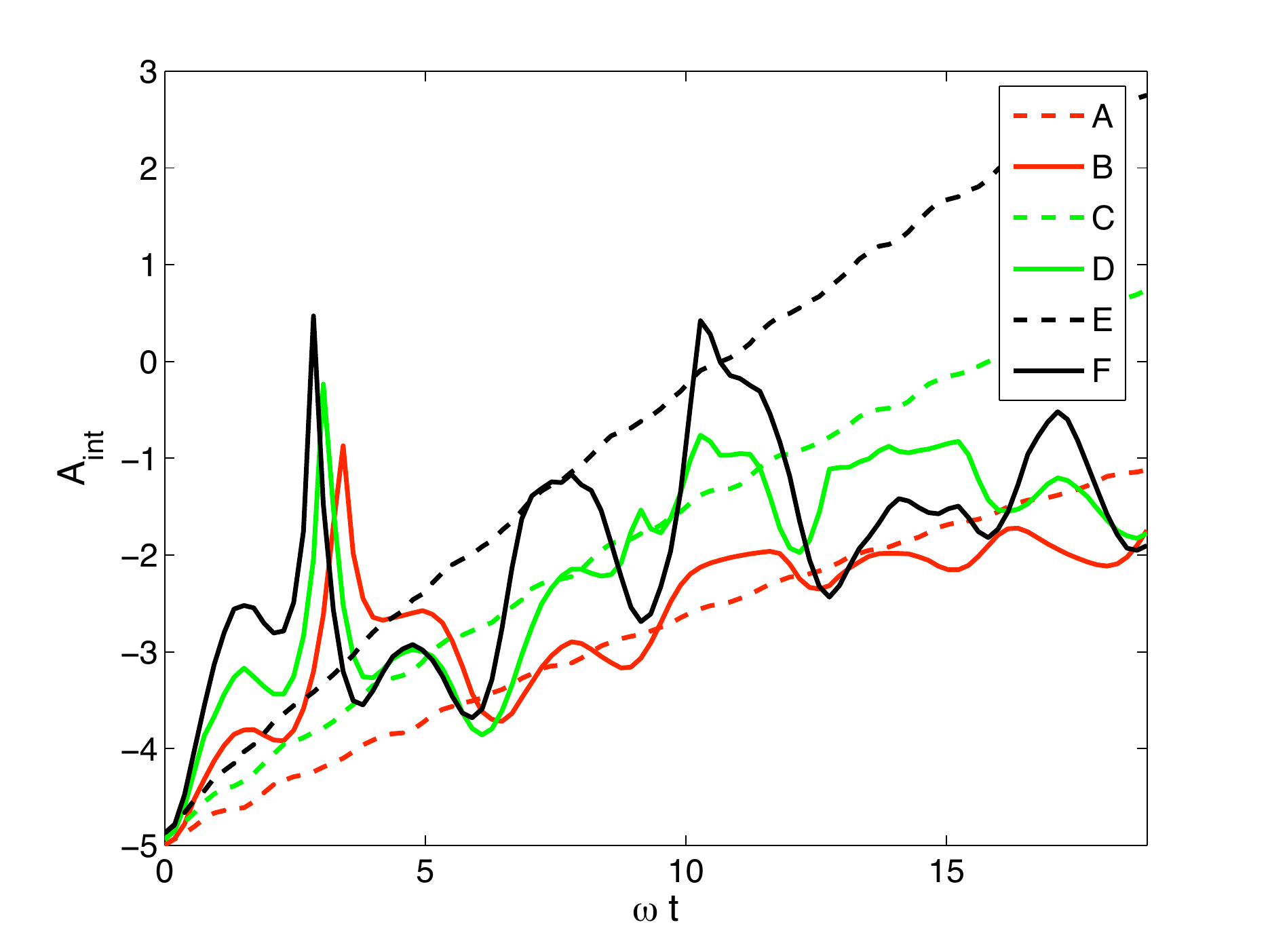}
	\caption{Displacement dependence of $A_{\mathrm{int}}$ vs. $\omega t$ for the resonator coupled to the  independent ohmic oscillator bath (A, C, and E) and to the three TLS's with energies indicated in Fig.~\ref{quadrant} without the oscillator's ohmic bath (B,D, and F). The initial displacements are $x_0=5$  (A, B), $x_0=6$ (C, D), and $x_0=7$ (E, F). The parameter values are   $\lambda=0.1$, $\;\Delta_0=0,\;\Delta_b=1,\;T=1,\;\mathrm{and}\; T_1=10.$  }
	\label{A_intOscSpin}
\end{figure}

\begin{figure}[htbp]
	\centering
		\includegraphics[height=3in]{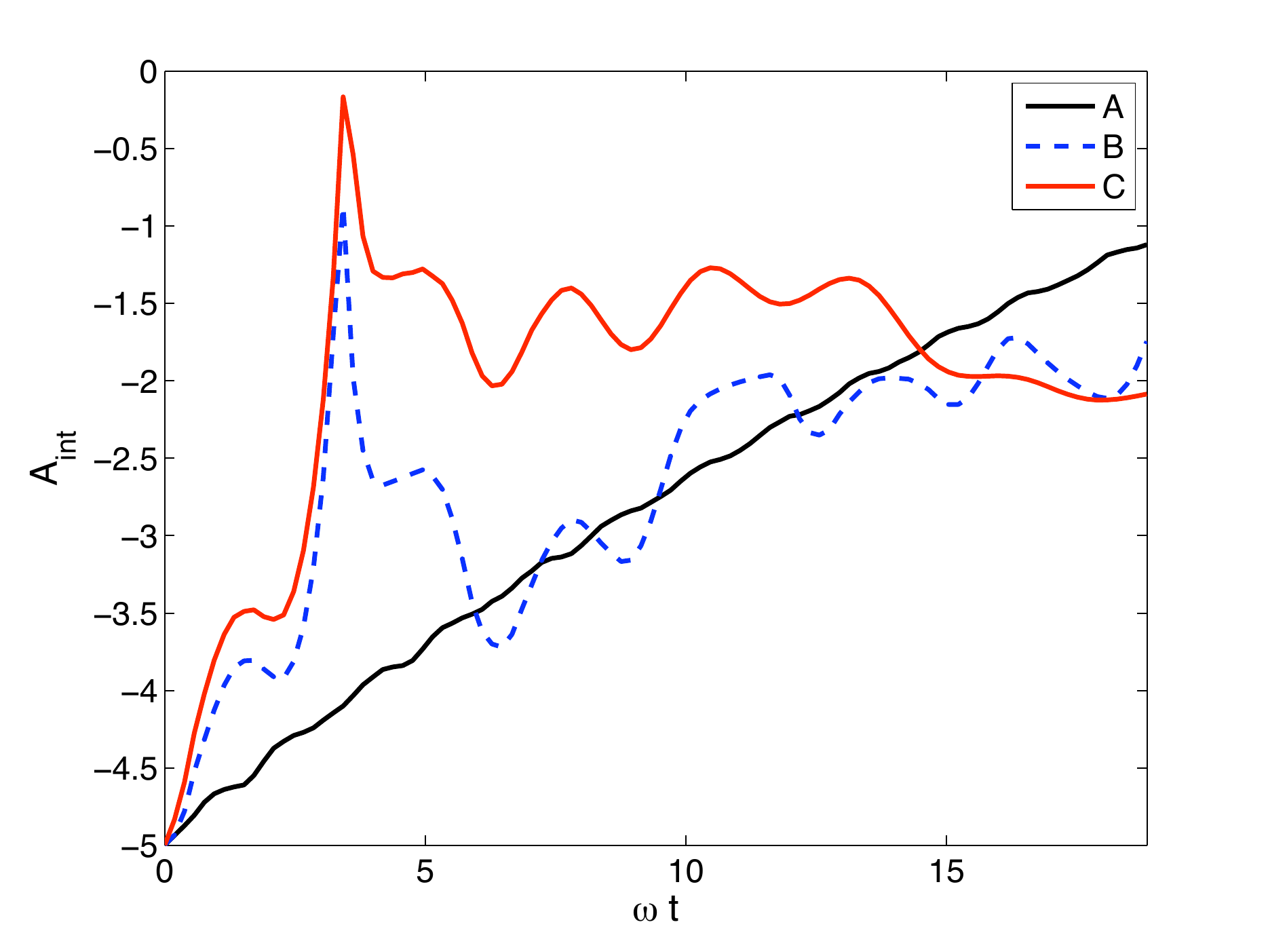}
	\caption{$A_{\mathrm{int}}$ vs. $\omega t$ for the resonator coupled to the independent ohmic oscillator bath only (A), to three TLS's only (B), and to the  ohmic oscillator bath and the three TLS's (C).  The TLS $\Delta_i$ values are the same as those  in Fig.~\ref{A_intOscSpin}.  The initial displacement is $x_0=5$, and the parameter values are $\lambda=0.1,\;\gamma=0.01,\;T=1,\;\mathrm{and}\; T_1=10.$}
	\label{A_intOsc3Spins}
\end{figure}

In Fig.~\ref{A_intOscSpin} we plot the decoherence of the resonator coupled to three TLS's for three initial displacements (A, C, and E), as well as the decoherence of the resonator coupled only to the oscillator bath for comparison.  The decoherence due to the presence of the three TLS's is greater than that due to a single TLS, while the decoherence due to the oscillator bath shows a systematic dependence on displacement that the TLS-induced decoherence does not exhibit. Decoherences and partial recoherences on the shorter, mechanical period timescales can be clearly seen,  due to the mechanical resonator-TLS's entangling-disentangling dynamics  as discussed above for the single TLS case.  Similar behavior was found for other randomly selected sets of three TLS's with energies near the resonator energy. Finally, in Fig.~\ref{A_intOsc3Spins} we plot $A_{\mathrm{int}}$ vs. $\omega t$ for the resonator coupled only to the ohmic bath (A), coupled only to the three TLS's  (B), and coupled to both the ohmic bath and the three TLS's (C).  As expected, we find that decoherence occurs more rapidly when the oscillator is coupled to both the ohmic bath and to the three TLS's.

\section{\label{sec:conclusion}Conclusion}      
In the present paper we have investigated the damping and decoherence dynamics of the flexural mode of a nanomechanical beam resonator interacting with a few damped TLS's. It was found that the resulting damping rate is amplitude dependent, while the decoherence rate for superpositions of position states depends only weakly on their separation. This is to be contrasted with the  damping and decoherence trends of the more commonly considered  resonator  interacting bilinearly with an  ohmic bath of free oscillators. In the latter case, the resulting damping is amplitude-independent, while the decoherence rate scales with the square of the position separation in the initial superposition state. 

In our model, strain-mediated interactions between the TLS's were neglected. It will be interesting to take such interactions into account; the reduced volume of a suspended, nanoscale resonator may result in a significant enhancement of the strain-mediated TLS-TLS interaction as compared with  the TLS-TLS interaction in the  bulk. It is also important to try to increase the number of TLS's in our numerical experiments and compare the obtained damping and decoherence trends with  approximate analytic results derived for the model system of a resonator interacting weakly with a dense spectrum of TLS's.  This will then enable a test of the  analytical approximations, as well as help shed light on the role of TLS's in the larger, optomechanical resonator experiments.

\section*{Acknowledgements}
We thank Andrew Armour  for helpful discussions and for reading the manuscript. This work was partly supported by the National Science Foundation under Grant  Nos. CMS-0404031 and DMR-0804477 (L.R. and M.B.), the Foundational Questions Institute (L.R. and M.B.), and by the Japanese Society for the Promotion of Science (M.B. and Y.T.). 

\appendix
\section{\label{sec:masterequation}Derivation of the Oscillator-TLS dissipative master equation}
In terms of the system-environment density matrix $\rho$ and  Liouvillian superoperator $L$ defined by $L\rho =[H,\rho]$ where $H$ is the total system-environment Hamiltonian, the time-dependent Schr\"{o}dinger equation is 
\begin{equation}
\dot{\rho}=-\frac{i}{\hbar}L\rho.
\label{schreq}
\end{equation} 
We wish to derive a master equation for the system density matrix comprising the oscillator and TLS only: $\rho_S ={\mathrm{Tr}}_{\mathrm{env}}\rho$, where the trace is performed over the oscillator-TLS environment. Following the Nakajima-Zwanzig projection operator method\cite{nakajima58,zwanzig60} along the lines of Ref.~[\onlinecite{divincenzo05}], we introduce projectors
\begin{equation}
P(\cdot)=\rho_{\mathrm{env}} \otimes{\mathrm{Tr}}_{\mathrm{env}} (\cdot)
\label{Pprojectoreq}
\end{equation}
and
\begin{equation}
Q(\cdot)=(1-\rho_{\mathrm{env}}\otimes {\mathrm{Tr}}_{\mathrm{env}})(\cdot),
\label{Qprojectoreq}
\end{equation}
where  $\rho_{\mathrm{env}}=e^{-\beta H_{\mathrm{env}}}/Z$, with $\beta=1/k_B T$, for environment temperature $T$. Suppose that at the initial time $t=0$, the system and environment are in a product state $\rho(0)=\rho_S(0)\otimes\rho_{\mathrm{env}}$. Then $P\rho(0)=\rho(0)$ and $Q\rho(0)=0$. Now partition the density matrix $\rho$ and Schr\"{o}dinger equation using the projectors $P$ and $Q$: 
\begin{eqnarray}
\dot{\rho}_1 &=&-\frac{i}{\hbar} P L(\rho_1 +\rho_2)\label{projectedmastereq1}\\
\dot{\rho}_2 &=&-\frac{i}{\hbar} (1-P) L(\rho_1 +\rho_2),
\label{projectedmastereq2}
\end{eqnarray}
where $\rho_1 =P\rho$ and $\rho_2=Q\rho$. Solving formally for $\rho_2$, we have
\begin{equation}
\rho_2(t)=-\frac{i}{\hbar}\int_0^t d t' \exp\left[-\frac{it'}{\hbar} (1-P) L\right] (1-P)L\rho_1(t-t'),
\label{rho2solneq}
\end{equation}
where we have used the fact that $\rho_2(0)=Q\rho(0)=0$.
Substituting Eq.~(\ref{rho2solneq}) into Eq.~(\ref{projectedmastereq1}), we obtain
\begin{equation}
\dot{\rho}_1=-\frac{i}{\hbar} PL\rho_1 -\frac{1}{\hbar^2}\int_0^t dt' PL\exp\left[-\frac{i}{\hbar}(t-t')(1-P) L\right] (1-P)L\rho_1 (t').
\label{rho1effectiveeq}
\end{equation}
Using the definition for the Liouvillian superoperator $L$ given above and assuming
\begin{equation}
{\mathrm{Tr}}_{\mathrm{env}}(b_n \rho_{\mathrm{env}})={\mathrm{Tr}}_{\mathrm{env}}(c^{(j)}_n \rho_{\mathrm{env}})=0, 
\label{annihileq}
\end{equation}
Eq.~(\ref{rho1effectiveeq}) simplfies to
\begin{equation}
\dot{\rho}_S=-\frac{i}{\hbar} L_S \rho_S -\frac{1}{\hbar^2}\int_0^t dt' \Sigma_S(t-t') \rho_S (t'),
\label{systemeffectiveeq}
\end{equation}
where the self-energy superoperator kernel is
\begin{equation}
\Sigma_S (t)={\mathrm{Tr}}_{\mathrm{env}}\left[L_{S-\mathrm{env}} \exp(-iQLt/\hbar) L_{S-\mathrm{env}} \rho_{\mathrm{env}}\right]
\label{kerneleq}
\end{equation} 
and where $L_S$ is the system part and $L_{S-{\mathrm{env}}}$ the system-environment part of the full Liouvillian superator $L$. 

We now make the Born approximation, which amounts to dropping the interaction part $L_{S-{\mathrm{env}}}$ from the full $L$ appearing in the exponential term of the kernel:
\begin{eqnarray}
\Sigma_S(t)&\approx &{\mathrm{Tr}}_{\mathrm{env}}\left[L_{S-\mathrm{env}} e^{-iQ(L_S +L_{\mathrm{env}})t/\hbar} L_{S-\mathrm{env}} \rho_{\mathrm{env}}\right]\cr
&=&{\mathrm{Tr}}_{\mathrm{env}}\left[L_{S-\mathrm{env}} e^{-i(L_S +L_{\mathrm{env}})t/\hbar} L_{S-\mathrm{env}} \rho_{\mathrm{env}}\right],
\label{bornapproxeq}
\end{eqnarray}
where we can drop the $Q$ projector as in the last line, a consequence of Eq.~(\ref{annihileq}). We assume that the system-environment interaction is sufficiently weak to justify making this Born approximation.

Given the bilinear in operators form of the system-environment interaction Hamiltonian $H_{S-{\mathrm{env}}}$ [third and fourth terms in Eq.~(\ref{envhamiltonianeq})] and using the  following identity for any two operators $A$ and $B$
\begin{equation}
{\mathrm{Tr}} (AB)=\frac{1}{2}\sum_{\nu=0}^3 {\mathrm{Tr}}(A\sigma_{\nu}) {\mathrm{Tr}}(B\sigma_{\nu}),
\label{sigmaidentityeq}
\end{equation} 
where $\sigma_i$, $i=1,2,3$ are the Pauli matrices and $\sigma_0$ is the identity matrix, the Born approximation to the master equation (\ref{systemeffectiveeq}) can be rewritten after some algebra as follows:   
\begin{eqnarray}
\dot{\rho}_S(t)&=& -\frac{i}{\hbar}[H_S,\rho_S(t)]-\frac{1}{\hbar^2}\int_0^t dt'{\mathrm{Tr}}_{\mathrm{env}} \left[B B(-t')\rho_{\mathrm{env}}\right]\left[Y,Y(-t')e^{-i H_S t'/\hbar }\rho_S (t-t') e^{+i H_S t'/\hbar}\right]\cr
&&-\frac{1}{\hbar^2}\sum_{j=1}^N\int_0^t dt'{\mathrm{Tr}}_{\mathrm{env}} \left[C^{(j)} C^{(j)}(-t')\rho_{\mathrm{env}}\right]\left[\sigma_z^{(j)},\sigma_z^{(j)}(-t')e^{-i H_S t'/\hbar }\rho_S (t-t') e^{+i H_S t'/\hbar}\right]\cr
&& + {\mathrm{h. c.}}.
\label{bornmastereq}
\end{eqnarray}
In this expression, $O(t) =e^{iH_S t/\hbar} O e^{-iH_St/\hbar}$ for some operator $O$, ``h.c." denotes the hermitian conjugate of the preceding integral terms, and  we have used the shorthand notations $B=\sum_n \kappa_n (b^+_n +b_n)$ and $C^{(j)}=\sum_n \tilde{\kappa}^{(j)}_n(c_n^{(j)+}+c_n^{(j)})$.  Using the fact that the environment is in a thermal state to work out the environment correlation functions, we obtain:
\begin{equation}
{\mathrm{Tr}}_{\mathrm{env}} \left[B B(t)\rho_{\mathrm{env}}\right]=\int_0^{\infty}d\omega J(\omega) \left[\coth(\beta\hbar\omega/2) \cos(\omega t) +i\sin(\omega t)\right],
\label{correlation1eq}
\end{equation}
where the bath spectral density is
\begin{equation}
J(\omega)=\sum_n \kappa_n^2 \delta(\omega -\omega_n),
\label{spectraldensityeq}
\end{equation}
and we have a similar equation for the correlation function involving $C^{(j)}$:
\begin{equation}
{\mathrm{Tr}}_{\mathrm{env}} \left[C^{(j)} C^{(j)}(t)\rho_{\mathrm{env}}\right]=\int_0^{\infty}d\omega \tilde{J}^{(j)}(\omega) \left[\coth(\beta\hbar\omega/2) \cos(\omega t) +i\sin(\omega t)\right],
\label{correlation2eq}
\end{equation}
with
\begin{equation}
\tilde{J}^{(j)}(\omega)=\sum_n (\tilde{\kappa}^{(i)}_n)^2 \delta(\omega -\omega_n).
\label{spectraldensity2eq}
\end{equation}
For simplicity, we assume an ``ohmic" spectral density with power law cutoff:
\begin{equation}
J(\omega)=\alpha \omega\frac{ \Lambda^2}{\omega^2 +\Lambda^2},
\label{ohmiceq}
\end{equation}
where $\Lambda$ is the ultraviolet cutoff frequency.
Substituting (\ref{ohmiceq}) into (\ref{correlation1eq}) and using contour integration to solve for the integral, we obtain:
\begin{equation}
{\mathrm{Tr}}_{\mathrm{env}} \left[B B(t)\rho_{\mathrm{env}}\right]=\frac{\pi\alpha\Lambda^2}{2}\left[e^{-\Lambda t} \cot(\beta\hbar\Lambda/2) +\frac{4}{\beta\hbar^2}\sum_{n=1}^{\infty}\frac{ \left({2\pi n}/{\beta}\right) e^{-2\pi n t/\beta\hbar}}{\left({2\pi n}/{\beta\hbar}\right)^2 -\Lambda^2} +i e^{-\Lambda t}\right],
\label{ohmiccorrelationeq}
\end{equation}
with a similar expression for  the correlation function (\ref{correlation2eq}).  

From the form of the environment correlation function (\ref{ohmiccorrelationeq}), we see that it decays rapidly to zero relative to the oscillator and TLS dynamical timescales, provided we assume that the  environment temperature satisfies  $k_B T\gg \hbar \omega, E^{(j)}$, where $E^{(j)}=\sqrt{(\Delta_0^{(j)})^2+(\Delta_b^{(j)})^2}$ is the $j$th TLS energy level separation. Subject to this condition on the temperature, we can make a Markov approximation in master equation (\ref{bornmastereq}) by setting $t\rightarrow \infty$ in the upper integration limit and expanding to first order in time the system's $t'$-dependent unitary  evolution operator wherever it appears, i.e.,   $e^{\pm i H_S t'}\approx 1\pm i H_S t'$.  The resulting Born-Markov  master equation is
\begin{eqnarray}
\dot{\rho}_S (t)&=&-\frac{i}{\hbar}[H_S,\rho_S(t)]\cr
&&+\frac{iI_1 }{\hbar^2} [Y^2,\rho_S(t)]-\frac{iI_2}{\hbar^2 m} [Y,\{ P_Y,\rho_S(t)\}]-\frac{ R_1}{\hbar^2}[Y,[Y,\rho_S(t)]]+\frac{ R_2}{\hbar^2 m} [Y,[P_Y,\rho_S (t)]]
\cr&&-\frac{1}{\hbar^2}\sum_{j=1}^N\tilde{R}^{(j)}_1[\sigma^{(j)}_z,[\sigma^{(j)}_z,\rho_S(t)]]-\frac{i}{\hbar^3}\sum_{j=1}^N\Delta^{(j)}_b \tilde{I}^{(j)}_2[\sigma^{(j)}_z,\{\sigma^{(j)}_y,\rho_S(t)\}]\cr
&&+\frac{1 }{\hbar^3} \sum_{j=1}^N\Delta_b^{(j)}\tilde{R}^{(j)}_2[\sigma^{(j)}_z,[\sigma^{(j)}_y,\rho_S(t)]],
\label{BornMarkovMastereq}
\end{eqnarray}
where $P_Y$ is the oscillator momentum, $\{\cdot,\cdot\}$ denotes the anticommutator, and the $R$ and $I$ coefficients are the real and imaginary parts of the environment correlation function time integrals:
\begin{equation}
R_1={\mathrm{Re}}\int_0^{\infty} dt  {\mathrm{Tr}}_{\mathrm{env}} \left[B B(t)\rho_{\mathrm{env}}\right],
\label{osccorreq}
\end{equation} 
\begin{equation}
R_2={\mathrm{Re}}\int_0^{\infty} dt  t{\mathrm{Tr}}_{\mathrm{env}} \left[B B(t)\rho_{\mathrm{env}}\right],
\label{osccorr2eq}
\end{equation} 
\begin{equation}
\tilde{R}^{(j)}_1={\mathrm{Re}}\int_0^{\infty} dt  {\mathrm{Tr}}_{\mathrm{env}} \left[C^{(j)} C^{(j)}(t)\rho_{\mathrm{env}}\right],
\label{spincorreq}
\end{equation} 
\begin{equation}
\tilde{R}^{(j)}_2={\mathrm{Re}}\int_0^{\infty} dt t {\mathrm{Tr}}_{\mathrm{env}} \left[C^{(j)} C^{(j)}(t)\rho_{\mathrm{env}}\right]
\label{spincorr2eq}
\end{equation} 
and analogously for the imaginary parts. In the Born-Markov master equation (\ref{BornMarkovMastereq}), the $I_1$ term renormalizes the frequency of the oscillator, while the $R_2$ term is the so-called `anomalous diffusion' contribution. We shall neglect both terms, justified because of  the assumed weak system-environment coupling and the above condition on the environment temperature. The remaining oscillator environment terms involving the $I_2$ and $R_1$ cause damping and thermal diffusion of the oscillator, respectively. It is convenient to parametrize $I_2$ in terms of the energy damping rate $\gamma$ of the oscillator in the absence of the TLS: $I_2 =\hbar m \gamma/2$. The diffusion coefficient  then becomes $R_1= m\gamma k_B T$, the expected form that follows from the fluctuation-dissipation theorem. The effect of the remaining three TLS environment terms are most straightforwardly understood by considering the  coupled moment equations of the three Pauli matrices in the absence of the oscillator. One finds for weak system-environment coupling that the $\tilde{R}^{(j)}_2$ term can be neglected, while the $\tilde{R}^{(j)}_1$ term causes damping/dephasing of the TLS and the $\tilde{I}^{(j)}_2$ diffusion term ensures that the moments decay to the thermal equilibrium state.
Again, it is convenient to parametrize  $\tilde{R}^{(j)}_1$  in terms of the relaxation time $T^{(j)}_1$ of the $j$th TLS excited   eigenstate in the absence of the oscillator. From the moment equations, we obtain $\tilde{R}^{(j)}_1={\hbar^2}({E^{(j)}}/{\Delta^{(j)}_b})^2{(4T_1^{(j)})^{-1}}$ and $\tilde{I}_2^{(j)}=\hbar/(2 k_B T) \tilde{R}^{(j)}_1$, as follows from the fluctuation-dissipation theorem.  In terms of these parametrizations, master equation (\ref{BornMarkovMastereq}) becomes
\begin{eqnarray}
\dot{\rho}_S (t)&=&-\frac{i}{\hbar}[H_S,\rho_S(t)]-\frac{i\gamma}{2\hbar} [Y,\{ P_Y,\rho_S(t)\}]-\frac{m\gamma k_B T}{\hbar^2}[Y,[Y,\rho_S(t)]]\cr
&&-\sum_{j=1}^N\frac{1}{4T_1^{(j)}}\left(\frac{E^{(j)}}{\Delta^{(j)}_b}\right)^2[\sigma^{(j)}_z,[\sigma^{(j)}_z,\rho_S(t)]]
\cr&&-\sum_{j=1}^N\frac{i}{8T_1^{(j)}}\left(\frac{\left(E^{(j)}\right)^2}{\Delta^{(j)}_b k_B T}\right)[\sigma^{(j)}_z,\{\sigma^{(j)}_y,\rho_S(t)\}],
\label{BornMarkovMaster2eq}
\end{eqnarray}
where we recognize in the first line the familiar  master equation for a quantum Brownian oscillator in the large temperature limit.\cite{zurek91}

While the master equation (\ref{BornMarkovMaster2eq}) is valid in the large temperature limit $k_B T\gg \hbar \omega, E^{(j)}$, it is desirable to investigate the system dynamics at low temperatures as well, such that $k_B T\lesssim \hbar \omega, \Delta^{(j)}$. In principle,  a more involved  analysis of Eq.~(\ref{bornmastereq}) with correlation relation expressions (\ref{ohmiccorrelationeq})  can yield a Markovian approximation that is valid at lower temperatures. However, a more direct way is simply to invoke the quantum version of the fluctuation-dissipation theorem, which for a Brownian oscillator amounts to making the following replacement in the diffusion term in (\ref{BornMarkovMaster2eq}):
\begin{equation}
\frac{\hbar\omega}{2k_B T}\rightarrow\tanh\left(\frac{\hbar\omega}{2 k_B T}\right).
\label{qbmeq}
\end{equation}
The resulting Born-Markov  master equation describing the quantum Brownian motion of the oscillator alone is valid for temperatures $k_B T\gg \gamma\hbar$. Given the assumed  weak system-environment coupling, i.e., a large quality factor $Q=\omega/\gamma\gg 1$, oscillator dynamics can now be investigated at low temperatures such that $k_B T\lesssim \hbar \omega$. Analogously, we can make the replacement 
\begin{equation}
\frac{E^{(j)}}{2k_B T}\rightarrow\tanh\left(\frac{E^{(j)}}{2 k_B T}\right)
\label{qTLSeq}
\end{equation}        
in the second TLS (diffusion) term of  the  master equation (\ref{BornMarkovMaster2eq}). The resulting moments for the three Pauli matrices in the absence of the oscillator then decay to the correct quantum thermal equilibrium state as required. We shall assume that making the replacements (\ref{qbmeq}) and (\ref{qTLSeq}) in (\ref{BornMarkovMaster2eq}) yield the Born-Markov master equation that is valid for temperatures $k_B T\gg \gamma \hbar,\hbar/T_1$, provided the interactions between the oscillator and TLS, as well as between the oscillator-TLS system and environment are weak, i.e., $\lambda^{(j)}\ll \hbar\omega, E^{(j)}$; $Q\gg 1$ and $E^{(j)} T_1/\hbar\gg 1$.

\end{document}